\documentclass[aps,prd,twocolumn,amsfonts,amssymb,amsmath,showpacs,letterpaper,superscriptaddress,nofootinbib,altaffilletter,
fontenc]{revtex4-1}
\usepackage{amsmath,amssymb,graphicx}
\usepackage[english]{babel} 
\usepackage[printwatermark]{xwatermark}
\usepackage{xcolor}
\usepackage{lipsum}
\usepackage{graphicx}
\usepackage{multirow}
\usepackage{booktabs}
\usepackage{bm}
\usepackage{multirow}
\usepackage{color}
\usepackage{booktabs}
\usepackage{tabularx}
\usepackage{times}
\usepackage{microtype}
\usepackage{booktabs}
\usepackage{subfigure}
\usepackage[normalem]{ulem}
\usepackage[varg]{txfonts}
\usepackage{dirtytalk}
\usepackage[colorlinks, pdfborder={0 0 0}]{hyperref}
\definecolor{LinkColor}{rgb}{0.75 , 0, 0}
\definecolor{CiteColor}{rgb}{0, 0.5, 0.5}
\definecolor{UrlColor}{rgb}{0, 0, 0.75}
\hypersetup{linkcolor=LinkColor}
\hypersetup{citecolor=CiteColor}
\hypersetup{urlcolor=UrlColor}
\usepackage[utf8]{inputenc}
\usepackage{csquotes}
\usepackage[normalem]{ulem}

\colorlet{Mycolor1}{green!10!orange!90!}
\definecolor{Mycolor2}{HTML}{00F9DE}

\allowdisplaybreaks
\newcommand{\FigStart}{\begin{figure*}[h]}
\newcommand{\FigEnd}{\end{figure*}}

\newcommand{\fisco}{f_\text{ISCO}}

\def\imrPhenomPv2{\texttt{IMRPhenomPv2}}
\def\taylorF2{\texttt{TaylorF2}}

\begin{document}
\title{Testing the nature of compact objects in the lower mass gap using gravitational wave observations} 
\author{N. V. Krishnendu}
\email{k.naderivarium@bham.ac.uk}
\affiliation{School of Physics and Astronomy, University of Birmingham, Edgbaston, Birmingham, B15 2TT, UK}
\affiliation{International Centre for Theoretical Sciences (ICTS), Survey No. 151, Shivakote, Hesaraghatta, Uttarahalli, Bengaluru, 560089, India}
\author{Frank Ohme}
\email{frank.ohme@aei.mpg.de}
\affiliation{Max Planck Institute for Gravitational Physics (AEI), Callinstraße 38, Hanover, 30167, Germany}
\author{K. G. Arun}
\email{kgarun@cmi.ac.in}
\affiliation{Chennai Mathematical Institute, Siruseri, 603103, India}
\date{\today}
\begin{abstract}
As the compact binary catalog continues to grow rapidly, developing and refining tests to probe the nature of compact objects is essential for a comprehensive understanding of both the observed data and the underlying astrophysics of the binary population. We investigate the effectiveness of spin-induced multipole moments (SIQM) and tidal deformability measurements in distinguishing lower mass-gap black hole (BH) binaries from non-BH binaries with different mass and spin configurations. We perform model-agnostic tests on binary BH (BBH) simulations using full Bayesian inference, evaluating the independent and joint measurability of SIQM and tidal parameters across the parameter space. We extend the analysis to simulations of self-interacting spinning boson stars, using synthetic signals that exhibit (a) both SIQM and tidal effects and (b)  each effect individually. For case (a), recovery is performed using (i) a BBH model, (ii) a model incorporating both SIQM and tidal effects, and (iii) models including either SIQM or tidal effects. For case (b), we employ (i) a BBH model and (ii) models incorporating either SIQM or tidal effects,  consistent with the injection. Simulations assume binaries of total mass of $\rm{8 M_{\odot}}$, varying mass ratios and spin magnitudes, using the inspiral-only {\texttt{TaylorF2}} waveform model. A three-detector network consists of Advanced LIGO and Virgo at design sensitivity is assumed, with results scalable to future detector sensitivities. We find that employing an incorrect model to analyze the signal can lead to severe biases in parameter inference. Notably, when analyzing a simulated binary boson star-like signal with component masses $\rm{(4, 4) \, M_{\odot}}$ using a BBH model, the system is incorrectly identified as having masses $\rm{(8, 2) \, M_{\odot}}$. In contrast, using the correct recovery model that includes both SIQM and tidal deformability effects, successfully recovers the true masses, highlighting the significance of waveform model accuracy in performing reliable distinguishability tests for compact objects in the low-mass gap.
\end{abstract}
\pacs{} \maketitle
\section{Introduction}\label{intro}
Across the first four observing runs of the LIGO, Virgo, and KAGRA gravitational wave (GW) detectors, a total of 218 compact binary coalescence observations have been reported~\cite{GWTC4,GWTC3-catalog,GWTC-2-catalog}. The fourth observing run (O4) is ongoing, and the public alert system reports more than a hundred events in the second and third parts of the fourth observing run alone~\cite{LIGO2024_gracedb}. Independent analyses (see, for instance, ~\cite{Nitz:2021uxj, Wadekar:2024zdq}) using the data from the first three observing runs are in broad agreement with the findings of \cite{GWTC3-catalog,GWTC-2-catalog}. The fifth observing run (O5), with LIGO at the design sensitivity, is expected to enable many more such detections~\cite{ligo, AdvLIGO, AdvVirgo}. Moreover, at the end of O5, the LIGO-India detector is expected to join the network~\cite{LIGO-India}. Simultaneously, research and development efforts for the third-generation gravitational-wave detectors, Einstein Telescope and Cosmic Explorer, as well as the space-based detector, the Laser Interferometer Space Antenna, are actively underway,  with the hope of being operational in the next decade~\cite{Evans:2023euw, Branchesi:2023mws, lisa}. 

One novel class of sources detected via GW observations consists of binaries containing a compact object with mass in the putative `low mass gap' of $\sim 3–5 \rm{M}_{\odot}$~\cite{GW190814, GW230529}, conjectured to arise due to the existence of an upper limit on the neutron star mass and a lower limit on the black hole (BH) mass set by stellar evolution processes (see, e.g. \cite{Gupta:2019nwj, Mahapatra:2025agb, Littenberg:2015tpa} for ways to populate the low mass gap through hierarchical mergers). Determining the BH nature of the low mass gap compact objects has profound implications for stellar evolution,  dynamics in dense clusters and nuclear physics and has started attracting attention~\cite{Cotturone:2025jlm}. In this paper, we approach the problem from a fundamental physics standpoint and ask how currently employed tests of BH nature can be used to gain insights about these compact objects, what biases are expected in carrying out those tests with the present and future GW detectors, and how these biases may be remedied.
\subsection{Tests of black hole nature using gravitational waves}
Since the GW signals from merging compact objects imprint unique signatures of the dynamics of the binary, we can extract information about their internal structure by performing various tests on the detected signals  \cite{Johnson-Mcdaniel:2018cdu, Krishnendu:2017shb, Datta:2020gem, Cardoso2016, CardosoTidal2017, Cardoso:2017cfl, BC06, FH08, Ghosh:2025wex}. These GW-based tests of the nature of compact objects rely on measuring specific physical effects that can distinguish BHs from non-BH compact objects. For instance, the spinning motion of the compact object can deform the compact object in specific ways and is a characteristic of the internal composition of the star~\cite{Poisson:1997ha}. These spin-induced deformations can induce prolate and oblate differences in the compact shape of the object  and can be identified as the difference in spin-induced multipole moment imprints on GWs \cite{MKAF16,Bohe:2015ana, Marsat2015,Marsat2014}. The leading order effect is the spin-induced quadrupole moments (SIQMs)  and is an efficient probe to distinguish spinning BHs from other exotic compact objects~\cite{Krishnendu:2017shb, Krishnendu:2019tjp, GWTC-2-TGR, GWTC-3-TGR}.  The SIQM parameter can be parametrized for a compact object with mass $m$ and dimensionless spin, $\boldsymbol{\chi}$, as $Q=-\kappa\,m^3\boldsymbol{\chi}^2$, where $\kappa_{\rm BH}=1$. Non-BH objects can have values of $\kappa$ significantly different from 1. For instance, for spinning boson stars, the values of the SIQM parameters vary between $\sim 10$ and $150$~\cite{Ryan97,Pacilio:2020jza}, while for neutron stars, they range between $\sim 2$ and $14$~\cite{PappasMultipole2012}. In contrast, proposed models of slowly spinning gravastars predict SIQM values that can even be negative~\cite{Uchikata2015}.

The second routinely employed test is based on the GW measurements of tidal deformability of compact objects. Tidal deformations refer to distortions induced in a compact object by an external gravitational field~\cite{Hinderer:2007mb,1PNTidal2011,Flanagan:2007ix,Johnson-Mcdaniel:2018cdu, Sennett:2017etc,Vines:2011ud,DNV2012}.  It is the measurement of tidal deformability that led to the first GW-based constraints on the nuclear equation of state of the neutron star from the first binary neutron star signal, GW170817~\cite{GW170817}. The value of the tidal deformability parameter is uniquely determined for BHs. In contrast,  for neutron stars and other exotic stars, the tidal deformability parameter is a function of the star's internal structure. The tidal deformability of the primary neutron star in GW170817 is constrained to be $\leq 239$, with a stringent spin prior, at a $90\% $ confidence interval. This bound on the tidal deformability parameter can directly translate into the equation of state of the neutron star and is used to rule out/in many proposed alternate models~\cite{GW170817}.

Currently, the GW-based tests of the nature of compact objects measure SIQM and tidal deformability effects independently of each other. However, there is no fundamental reason to treat these effects in isolation for astrophysical compact objects in binary systems, as the non-BH nature would likely manifest in both of these physical effects simultaneously. Further, achieving the most accurate self-consistent description of such systems requires incorporating all known physical effects~\cite{Pacilio:2020jza} such as spin-induced deformations~\cite{Poisson:1997ha, Krishnendu:2017shb}, tidal deformability~\cite{Cardoso:2017cfl, Vines:2011ud, Flanagan:2007ix, Sennett:2017etc, Johnson-Mcdaniel:2018cdu}, spin-tidal interactions~\cite{Castro:2022mpw, Abdelsalhin:2018reg}, and tidal heating~\cite{Alvi:2001mx}. Each effect becomes important at different stages of the binary dynamics of an inspiralling binary (such as early or late inspiral), and using more than one effect is particularly powerful if the detected signal originates from a non-BH binary. 

 However, it is almost impossible to measure all these parameters together due to parameter correlations and the increase in the dimensionality of the parameter space, even for an ideal binary with the largest signal-to-ratio, which is analysed using the most accurate waveform model available to date. Hence, the two dominant physical effects, the tidal deformability and SIQM effects, are considered individually in most studies~\cite{Krishnendu:2017shb, Hinderer:2007mb}. However, there have been attempts to combine the SIQM and tidal effects and measure them simultaneously using GWs. Reference~\cite{Narikawa:2021pak} introduced a Bayesian inference-based framework for simultaneously measuring SIQM effects and tidal-deformability parameters in a model-agnostic way, considering inspiral-dominated signals in the second GW transient catalogue. Our approach is similar in spirit to theirs, but focuses on the low-mass gap compact objects with non-BH features.

When it comes to specific non-BH models, BH mimickers such as boson stars, theoretical models predict that their masses can overlap with BHs and neutron stars~\cite{Cardoso:2017cfl}, can have spin magnitudes similar to neutron stars, or exceeding the Kerr limit (super spinars~\cite{Uchikata:2021jmy}, self-spinning boson stars~\cite{Ryan97b}). Therefore, developing methods to simultaneously measure multiple physical effects, particularly SIQM and tidal deformability effects, is essential for reliably probing the nature of such compact objects. There have been proposals to construct an effective single parameter that carries information about the spin deformations as well as tidal deformations and constraining the properties of the exotic nature of the star, focusing on self-interacting spinning boson stars~\cite{Vaglio:2022flq, Vaglio:2023lrd, Pacilio:2020jza}. The motivation for such proposals is to account for the complete physical picture that if a compact star is affected by a neighbouring gravitational potential (its binary companion) and simultaneously is also spinning, then the emitted GW signal will carry both SIQM effects and tidal deformations. Reference~\cite{Vaglio:2023lrd} implemented a phenomenological relation connecting the boson star fundamental parameter,  a combination of the boson mass and the coupling constant in the gravitational potential in this case, to the SIQM parameters and tidal deformations of the individual compact objects in the binary and discussed the possible constraints from GWs. Instead of measuring the SIQM parameters and tidal deformability parameters from the data, they discuss the measurement of the boson star fundamental parameter, which is used to get constraints on the boson star model parameters.
\subsection{Determining the black hole nature of low mass gap objects}
The detection and parameter inference of  compact binaries in the existing GW transient catalogue reveal that the majority of observations are consistent with nearly equal mass, slowly spinning binaries composed of BHs and neutron stars~\cite{GWTC-2.1-catalog,GWTC4,GWTC3-catalog,GWTC-2-catalog}. Nevertheless, the detected events also include a  subpopulation of exceptional events with large mass asymmetry, significant spin effects, eccentricity signatures and compact objects in lower or higher mass gaps~\cite{GW190412,GW190425,GW190521,GW190814, Hannam:2021pit, Morras:2025xfu}. 

The detection GW230529~\cite{GW230523}, a binary with primary and secondary masses in the range $(2.5$-$4.5\rm{M}_\odot)$ and $(1.2$-$2\rm{M}_\odot)$ within the $90\%$ credibility, was widely discussed due to its unconventional properties. GW230529 was the first binary signal composed of a low-mass-gap BH paired with a neutron star. Since the primary mass falls below the typical BH mass range, the nature of the primary in GW230529 has been investigated extensively~\cite{Chandra:2024ila}. Moreover, many astrophysical formation and evolution scenarios have been proposed to explain the dynamical history of such an exceptional event~\cite{Janquart:2024ztv, vanPutten:2024ftm, Ye:2024wqj, Matur:2024nwi, Chatziioannou:2024tjq, Julie:2024fwy, Gao:2024rel, Datta:2020gem, Bhattacharya:2025xko}. The non-detection of associated electromagnetic counterparts also made it difficult to learn more about the astrophysical environment in which it formed~\cite{vanPutten:2024ftm}. In addition, a dedicated set of analyses was carried out to perform various fundamental tests of general relativity~\cite{Sanger:2024axs}. Notably, GW230529 provides the most stringent constraints on Einstein-Gauss-Bonnet gravity to date~\cite{Sanger:2024axs}, outperforming previous BBH results~\cite{GWTC-2-TGR, GWTC-3-TGR}, due to the long and loud inspiral in the detector network. In summary, performing parameter inference and extracting astrophysical properties for such exceptional events is extremely challenging, yet crucial for understanding astrophysical population inference of the detected signals.

Determining the nature of compact objects with masses in the lower mass gap is especially challenging because it is often hard to rule out neutron stars with exceptionally large masses from low-mass BHs, opening room for alternative explanations~\cite{PBH190814, Tews:2020ylw,Fasano:2020eum,Zhang:2020zsc,Chen:2020fzm,LIGOScientific:2024elc}, from compact objects like boson stars~\cite{BosonStars}, stars with primordial origin~\cite{PBH190814}, gravastars~\cite{Uchikata2015,Uchikata2016} and other more exotic models~\cite{CardosoTidal2017}. This implies that the GW measurements of SIQM and tidal deformability estimates could be a fine distinguisher in this regime, but several exotic alternatives can have the same mass and spin. One has to focus on the SIQM and tidal deformability  measurements and perform the most reliable distinguishability test. These tests are often limited by the unavailability of GW waveform models that describe the full inspiral-merger-ringdown regions of the compact binary dynamics, incorporating the extra matter effects induced by SIQM effects and tidal deformations. Additionally, these tests are limited by the sensitivity of current detectors in extracting the binary properties accurately, where the dominant noise systematic and noise characterization uncertainties can lead to false inference on the SIQM and tidal deformability estimates. It is also important to note that the effectiveness of these tests depends on the intrinsic properties of the binary. For instance, constraints based on SIQMs are most effective for binaries with large spin magnitudes~\cite{Krishnendu:2019tjp}. 
\subsection{This work}
 In this work, we focus on compact binaries for which both of the components lie in the lower mass gap. To understand and concretely address the issues related to their inference and confident identification, we study the feasibility of performing SIQM and tidal deformability–based tests of BH nature on simulated binaries and quantify the possible systematic biases that could arise due to employing incorrect model for the analysis.  We consider binaries of masses $\rm{(4+4)M_{\odot}}$ and $\rm{(5+3)M_{\odot}}$,  four different spin magnitudes and spin directions fixed along the direction of the orbital angular momentum axis. The details of the simulation setup are provided below.
\begin{enumerate}
    \item  We inject BBH signals and analyse them using a (a) BBH waveform, (b) parametrized SIQM waveforms and (c) parametrized tidal deformability waveforms. We compare the recovered binary parameters in these three cases and verify the ability to perform null tests for BBHs in the lower mass gap with current detectors employing SIQM and tidal deformability effects. When analyzing a BBH signal using recovery models (a) and (b), biases are expected to arise due to correlations between the binary parameters and the SIQM or tidal deformability parameters. We aim to quantify the magnitude of these biases and identify the regions of parameter space where they are most significant.
    \item Second, we create mock-binary signals with boson star signatures with SIQM and tidal deformability parameters. These simulations are then analysed assuming a (a) boson star model and a (b) BBH model. The binary boson star model is expected to provide the correct inference, and the inference from the binary BH (BBH) model can be used to quantify the systematic biases in binary parameters that arise due to the employment of an incorrect model.
    \item Third, to test the limitations of parameter recovery that uses only one physical effect, we generate mock boson star simulations that include only SIQM effects and analyse using the (a) BBH model and (b) SIQM model, similarly, for tidal deformability effects. Though this is not a physically motivated scenario, as the exotic compact object is expected to have non-BH features due to spin-induced deformations and tidally-induced deformations, our motivation here is to study the possible systematic biases in routinely employed methods for tests of BH nature.
\end{enumerate}
Table~\ref{tab:injection_recovery} summarizes these three cases.

This paper is organized in the following way. In Sec.~\ref{sec:details}, we provide the details about the waveform model and the Bayesian inference-based parameter estimation methods employed. Following that, Sec.~\ref{sec:results} details our findings. Specifically, we discuss the analysis performed on BBH simulations in ~\ref{subsec:bbh} and binary BSs in ~\ref{subsec:bosonstar} and in ~\ref{subsec:bosonstar2}. Finally, we conclude with Sec.~\ref {sec:conclusion}, highlighting the caveats and future prospects.
\vskip 0.5cm
\begin{table*}[htbp]
    \centering
    \renewcommand{\arraystretch}{1.8}
    \setlength{\tabcolsep}{6pt}
    \begin{tabular}{|l l l|}
        \hline
        \textbf{Injection model} & \textbf{Recovery model} & \textbf{Label} \\
        \hline

        \multirow{3}{*}{\parbox{5cm}{\raggedright BBH simulations: \\ $\{\delta\kappa_1=0, \delta\kappa_2=0\}$ \\ and $\{\Lambda_1=0, \Lambda_2=0\}$}} 
        & \parbox{5cm}{(i) BBHs} & $\rm{(BBH)_{inj} : (BBH)_{rec}}$ \\
        & \parbox{5cm}{(ii) Recovery includes \\SIQM parameters $\{\delta\kappa_1, \delta\kappa_2\}$} & $\rm{(BBH)_{inj} : (SIQM)_{rec}}$  \\ 
        & \parbox{5cm}{\vspace{0.8em}(iii) Recovery includes \\ tidal-deformability \\ parameters $\{\Lambda_1, \Lambda_2\}$\vspace{0.8em}} & $\rm{(BBH)_{inj} : (tides)_{rec}}$  \\
        \hline

        \multirow{4}{*}{\parbox{5cm}{\raggedright Non-BBH simulations with \\ 
$\delta\kappa_1 = \delta\kappa_2 = \delta\kappa^{\rm{BS}} = 10$ \\ 
and \\ 
$\Lambda_1 = \Lambda_2 = \Lambda^{\rm{BS}} = 289$}} 
    & \parbox{5cm}{(i) BBHs} 
    & $\rm{(SIQM\_tides)_{inj} : (BBH)_{rec}}$ \\

    & \parbox{5cm}{\vspace{0.8em}(ii) Recovery includes \\ SIQM parameters $\{\delta\kappa_1, \delta\kappa_2\}$} 
    & $\rm{(SIQM\_tides)_{inj} : (SIQM)_{rec}}$ \\

    & \parbox{5cm}{\vspace{0.8em}(iii) Recovery includes \\ tidal-deformability parameters $\{\Lambda_1, \Lambda_2\}$} 
    & $\rm{(SIQM\_tides)_{inj} : (tides)_{rec}}$ \\

    & \parbox{5cm}{\vspace{0.8em}(iv) Recovery includes \\ both SIQM and tidal-deformability \\ parameters $\{\delta\kappa_1, \delta\kappa_2\}$ and $\{\Lambda_1, \Lambda_2\}$\vspace{0.8em}} 
    & $\rm{(SIQM\_tides)_{inj} : (SIQM\_tides)_{rec}}$ \\
\hline

        \multirow{3}{*}{\parbox{5cm}{\raggedright Non-BBHs with \\ $\delta\kappa_1=\delta\kappa_2=\delta\kappa^{\rm{BS}}=10$ \\ or \\ $\Lambda_1=\Lambda_2=\Lambda^{\rm{BS}}=289$}} 
        & \parbox{5cm}{(i) BBHs}  & $\rm{(SIQM)_{inj} : (BBH)_{rec}}$ and $\rm{(tides)_{inj}:(BBH)_{rec}}$ \\
        & \parbox{5cm}{(ii) Recovery includes $\{\delta\kappa_1, \delta\kappa_2\}$ }& $\rm{(SIQM)_{inj} : (SIQM)_{rec}}$ \\
        & \parbox{5cm}{(iii) Recovery includes $\{\Lambda_1, \Lambda_2\}$} & $\rm{(tides)_{inj} : (tides)_{rec}}$ \\
        \hline
    \end{tabular}
    \caption{Summary of various models used for injections and recovery in this work.}
    \label{tab:injection_recovery}
\end{table*}

\section{Waveform model and parameter estimation} \label{sec:details}
The following subsections summarize the waveform models employed and the parameter estimation settings for the present work.
\subsection{Waveform model}\label{sec_wf}
GW waveforms from non-precessing compact binaries in their inspiral phase can be accurately modelled using post-Newtonian (PN) formalism~\cite{Marsat:2014xea, BFH2012, ABFO08, K95, MBFB2012, BMFB2012, Bohe:2013cla, M3B2013, Bohe:2015ana, Mishra:2016whh}. Reference~\cite{Mishra:2016whh} computed spin-dependent terms in the phase and the amplitude of the \rm{TaylorF2} approximant \cite{BIOPS2009}. This includes quadratic-in-spin terms up to 3 PN order, the cubic-in-spin terms at the leading order (3.5 PN), and the spin-orbit effects up to the 4 PN order. See Eqn. 2.1 in ~\cite{Krishnendu:2018nqa} for more details. Similarly, there have been efforts to model the effects of tidal deformability of stars on the emitted GWs \cite{1PNTidal2011, DNV2012, Lackey:2013axa, Pannarale:2015jka, Agathos:2015uaa} and the \rm{TaylorF2} approximant includes the leading (5 PN) plus sub-leading order (6 PN) tidal-deformability effects. 

The schematic representation of GW waveform in frequency domain is,
\begin{equation}
\tilde{h}(f)= \mathcal{C}\,\mathcal{A}(f)\, e^{i \psi(f)}~,
\label{eq:wf}
\end{equation}

where $\mathcal{C}$ is an overall constant, $\mathcal{A}(f)$ is the amplitude and $\psi(f)$ is the phase.

Following the PN approximation for inspiralling compact binaries, the stationary phase approximation~\cite{DIS00,DIS01,DIS02} provides,
\begin{equation}
    \psi(f)=2\pi f t_c+\phi_c+\frac{3}{128 \eta v^5}\left(1+\psi^{\rm{PP}}+\psi^{\rm{SIQM}}+\psi^{\rm{tidal}}\right),
    \label{eq:phase}
\end{equation}
where $\eta=\frac{m_1 m_2}{(m_1+m_2)^2}$ is the symmetric mass ratio and is a function of masses $m_1$ and $m_2$. Furthermore, $t_c$ and $\phi_c$ are the time and phase at coalescence.  We include the 3.5 PN corrections to the point particle term,  $\psi^{\rm{PP}}$. Schematically, the 3.5 PN point particle phase corrections are of the form,
 \begin{equation}
    \psi^{\rm{PP}}=\sum_{k=0}^{n=7} \left(\phi_k v^k + \phi_{kl}v^k \ln v \right),
\end{equation}
the power of the PN expansion parameter $v=\left(\pi m f_{gw}\right)^{1/3}$ determines the PN order ($k=n/2$ is the $n^{th}$ PN order) and the coefficients $\phi_k$ are functions of binary mass and spin. The PN coefficients, which appear with logarithmic PN parameter dependence, are written separately. For the 3.5 PN accurate expression, the log terms appear at  2.5PN and 3PN orders. The  $\psi^{\rm{SIQM}}$ term in Eq.~\ref{eq:phase} includes corrections due to the spin-induced quadrupole moment parameters at 2 PN and 3 PN orders~\cite{Krishnendu:2018nqa, Krishnendu:2017shb}. 
\begin{widetext}
\begin{subequations}
\begin{eqnarray}
\label{eq:P4}
\psi^{\rm{SIQM}}_{\rm{2\,PN}} &=&
-\frac{5}{8}\,\boldsymbol{\chi}_\mathrm{s}^2\,
\Bigl[1+156 \,\eta +80 \,\delta  \,\kappa _a+80 (1-2 \,\eta ) \kappa _s\Bigr]
+\boldsymbol{\chi}_\mathrm{a}^2
\left[-\frac{5}{8}-50 \,\delta  \,\kappa _a-50 \kappa _s+100 \,\eta  \left(1+\kappa _s\right)\right]
\nonumber\\
&-&\frac{5}{4}\,\boldsymbol{\chi}_\mathrm{a}\,\boldsymbol{\chi}_\mathrm{s} 
 \Bigl[\delta +80\,(1-2 \,\eta )\,\kappa _a+80 \,\delta  \,\kappa _s\Bigr]\,,\\
\psi^{\rm{SIQM}}_{\rm{3\,PN}} &=&
\pi\,\Biggl[\frac{2270}{3}\,\delta
\,\boldsymbol{\chi}_\mathrm{a}
+\left(\frac{2270}{3}-520\,\eta\right)
\boldsymbol{\chi}_\mathrm{s}\Biggr]
+\boldsymbol{\chi}_\mathrm{s}^2
\left[-\frac{1344475}{2016}+\frac{829705}{504}\,\eta+\frac{3415}{9}\,\eta ^2
+\delta \left(\frac{26015}{28}-\frac{1495}{6}\,\eta\right) \kappa_a
\right.\nonumber\\&+&\left.
\left(\frac{26015}{28}-\frac{44255}{21} \,\eta -240 \,\eta ^2\right) \kappa _s\right]
+\boldsymbol{\chi}_\mathrm{a}^2
\left[-\frac{1344475}{2016}+\frac{267815}{252} \,\eta -240 \,\eta ^2+\delta 
   \left(\frac{26015}{28}
-\frac{1495}{6} \,\eta \right) \kappa
   _a+\left(\frac{26015}{28}
   \right.\right.\nonumber\\&-&\left.\left.
   \frac{44255}{21} \,\eta -240 \,\eta ^2\right) \kappa _s\right]
+\boldsymbol{\chi}_\mathrm{a}\,\boldsymbol{\chi}_\mathrm{s} 
 \left[\left(\frac{26015}{14}-\frac{88510}{21} \,\eta -480 \,\eta ^2\right) \kappa _a+\delta\, 
   \biggl[-\frac{1344475}{1008}+\frac{745}{18} \,\eta +\left(\frac{26015}{14}
   \right.\right.\nonumber\\&-&\left.\left.
   \frac{1495
   }{3} \,\eta\right) \kappa _s\biggr]\right]\,,
\label{eq:P6}
\label{eq:P4}
\end{eqnarray}
\end{subequations}
\end{widetext}
 In the above equation, the $\kappa_s=0.5\left(\kappa_1+\kappa_2\right)$ and $\kappa_a=0.5\left(\kappa_1-\kappa_2\right)$, where $\kappa_1$ and $\kappa_2$ are the individual spin-induced quadrupole moment parameters. $\delta=m_1-m_2/M$, with $M$ being the total mass of the binary is called the difference mass ratio. The terms $\boldsymbol{\chi}_\mathrm{a}$ and $\boldsymbol{\chi}_\mathrm{s}$ denote the dimensionless spin angular momentum unit vectors pointing to the orbital angular momentum axis and $\boldsymbol{\chi}_\mathrm{s}$ and $\boldsymbol{\chi}_\mathrm{a}$ are the symmetric and anti-symmetric combinations of dimensionless spin angular vectors, $\boldsymbol{\chi}_\mathrm{1}$ and $\boldsymbol{\chi}_\mathrm{2}$ respectively. 
 
We introduce parametric deviations of the form $\kappa_1=1+\delta\kappa_1$ and $\kappa_2=1+\delta\kappa_2$ in the waveform model. For Kerr BHs,  $\delta\kappa_1=\delta\kappa_2=0$, uniquely determined by the No-Hair conjecture given the BH mass and spin~\cite{Carter71,Hansen74}. The GW measurement of fractional deviations $\delta\kappa_1$ and $\delta\kappa_2$ provides constraints on the nature of the compact object. Any deviation from the BH value hints to non-BH signature in the data. 

 In Eq.~\ref{eq:phase}, the term $\psi^{\rm{tidal}}$ includes the tidal deformability information in the GW phase.
\begin{equation}
    \psi^{\rm{tidal}}=-\frac{39}{2}\tilde{\Lambda}v^{10}+\left( \frac{6595}{264}\delta\tilde{\Lambda}-\frac{3115}{64}\tilde{\Lambda}\right)v^{12}
    \label{eq:tides}
\end{equation}
The leading order term $\tilde{\Lambda}$ at 5 PN and the subleading order term $\delta\tilde{\Lambda}$ are functions of masses and tidal deformabilities of binary constituents, $m_1$, $m_2$, and $\Lambda_1$, $\Lambda_2$~\cite{Flanagan:2007ix,Hinderer:2007mb,Vines:2011ud}.
 For BBHs $\tilde \Lambda=0$, hence is a powerful tool to distinguish BHs from other compact stars. This \texttt{TaylorF2} waveform model with SIQM and tidal-deformability effects are available in the GW data analysis toolkit, {\texttt LALSimulation}~\cite{lalsuite}. In this study, we use this waveform model to evaluate the posterior probablity distributions on each binary parameter, as described below.
\subsection{Parameter Estimation}
In the GW data analysis, we start with the data $\rm{d}$, which contains both noise and signal. Here, noise is a random process while the signal is modelled following a particular hypothesis $\rm{H}$ and is a function of the complete set of binary parameters $\theta$. The initial prior probability distribution, $\rm{p(\theta|H)}$, restricts the range of $\theta$. If we assume that the noise is Gaussian wide-sense stationary, the likelihood function takes the form, 
\begin{equation}
    \rm{p\left(d|H, \theta\right)\propto \exp{\left[-\frac{\left(d-h|d-h\right)}{2}\right]}}~,
    \label{eq:likelihood}
\end{equation}
where $\rm{d}$ and $\rm{h}$ are the frequency domain data and signal respectively and $\rm{(d-h|d-h)}$ is the noise-weighted inner product. Once we estimate the likelihood function and the prior distribution on each parameter is known, Baye's theorem provides the posterior probability distribution on each parameter as follows:
\begin{equation}
    \rm{p(\theta|H, d)=\frac{p(\theta|H)\, p(d|H, \theta)}{p(d|H)}}~,
    \label{eq:posterior}
\end{equation}
 where $p(\theta|H)$ denotes the prior on the parameter $\theta$ and $p(d|H, \theta)$ denotes the likelihood function.
In addition, Bayesian evidence $\rm{p(d|H)}$ is a measure of how much data support the hypothesis $\rm{H}$, and is obtained by marginalizing the likelihood over the full prior volume. 
\begin{equation}
    \mathcal{Z}=\rm{p(d|H)}=\int \rm{p(\theta|H)p(d|\theta, H) d\theta}~.
\end{equation}
To perform GW data inference and estimate posterior probability distributions and Bayesian evidence on different models, we modified the Bilby~\cite{bilby_paper, pbilby_paper} library and used the dynesty package~\cite{dynesty}, which is a stochastic sampling algorithm based on the nested sampling technique. We ensure that the likelihood evaluation is truncated at a frequency corresponding to the innermost stable circular frequency of a Kerr BH~\cite{Favata:2021vhw, Husa:2015iqa,Ori:2000zn}.
\subsection{Details of the simulation}

 The non-BH simulations assume a SIQM parameter value of $\delta\kappa_1=\delta\kappa_2=10$ and tidal deformability parameter value of $\Lambda_1=\Lambda_2=289$, following the boson star model provided in Ref.~\cite{Pacilio:2020jza}. These numbers are obtained by assuming a self-interacting spinning boson star configuration. For a boson star formed by bosons of mass $M_{B}$, the tidal deformability is a function of the boson star mass and the boson mass, $\Lambda \equiv \Lambda (M, M_B)$, however, the ratio has to be $M/ M_B \sim 0.061$ and therefore limits the $\Lambda$ to $\geq 289$, providing a lower boundary~\cite{Pacilio:2020jza}. The lower bound on $\Lambda$ sets a corresponding lower bound on the $\delta\kappa$ parameter through the relation that connects tidal deformability to SIQM parameters for a spinning boson star with a self-interacting potential. Assuming a maximum mass determined by the condition $M / M_B \sim 0.061$, the values $\delta\kappa_1 = \delta\kappa_2 = 10$ and $\Lambda_1 = \Lambda_2 = 289$ represent the minimal possible values. For any other parameter choices, these values will always be larger. Therefore, this choice results in the least biased estimate. The details of other binary parameters are in Table~\ref{tab:masses_spins_snr}. We focus on binaries in the lower mass gap to ensure inspiral dominance in the signal and analyzed with different spins and mass ratios. To avoid systematic biases from the post-inspiral part of the simulated signal, we truncate the analysis to the frequency corresponding to the innermost stable circular frequency of Kerr BHs~\cite{Favata:2021vhw, Husa:2015iqa,Ori:2000zn}. This frequency is a function of the component masses of the binary and spin magnitudes, and the values are reported in the last column of Table~\ref{tab:masses_spins_snr}. 

 The luminosity distance to the source is varied so that the signal-to-noise ratio (SNR) of the network, considering the twin LIGO detectors and the Virgo detector, is $50$ and $100$ for different simulations. The choice is made so that the SIQM parameters and the tidal-deformability parameters of the individual objects in the binary are measured well within the noise uncertainties of the  detectors, hence providing ample opportunities to perform systematic bias studies.  

The first injection model in Table~\ref{tab:injection_recovery}, a BBH model, is analysed with three recovery models. Namely, a BBH model with parameters $\rm{\{\theta_{BBH}\}}$, a SIQM model where we measure $\rm{\{\theta_{BBH}, \delta\kappa_i\}}$ and the tidal deformability model with parameters $\rm{\{\theta_{BBH}, \Lambda_i\}}$. For the second injection model, where we introduce boson star signatures in the binary signal by introducing non-BH values for SIQM and tidal deformability parameters, we consider four recovery models; the BBH model $\rm{\{\theta_{BBH}\}}$, the SIQM model $\rm{\{\theta_{BBH}, \delta\kappa_i\}}$, the tidal deformability model $\rm{\{\theta_{BBH}, \Lambda_i\}}$ and a model with SIQM and tidal deformability parameters, $\rm{\{\theta_{BBH}, \delta\kappa_i, \Lambda_i\}}$. The third category of simulated signals consider two types of injections. First a non-BBH model with non-BH SIQM values and a non-BBH model with non-BH tidal deformability parameters. Both these injections are analysed with a BBH model $\rm{\{\theta_{BBH}\}}$ and respective SIQM model $\rm{\{\theta_{BBH}, \delta\kappa_i\}}$  and tidal deformability model $\rm{\{\theta_{BBH}, \Lambda_i\}}$. The results obtained from this analysis are described in the next section. 
\begin{table*}[ht]
\centering
\renewcommand{\arraystretch}{1.5}
\begin{tabular}{|c c c c|}
\hline
\textbf{Masses ($\rm{M}_{\odot}$)} & \textbf{Spin combinations} & \textbf{\begin{tabular}{@{}c@{}} Network signal to \\ noise ratio \end{tabular}} & \textbf{$\fisco$ (Hz)} \\
\hline
$(m_1, m_2) = (5, 3)$ & 
\begin{tabular}{@{}l@{}}
$(\boldsymbol{\chi_1, \chi_2}) = (0.6, 0.5)$ \\
$(\boldsymbol{\chi_1, \chi_2}) = (0.3, 0.2)$ \\
$(\boldsymbol{\chi_1, \chi_2}) = (0.5, 0.05)$ \\
$(\boldsymbol{\chi_1, \chi_2}) = (0.6, 0.2)$
\end{tabular}
& 50 and 100 &
\begin{tabular}{@{}l@{}}
1645.97 \\
1325.07 \\
1438.93 \\
1562.56
\end{tabular} \\
\hline
$(m_1, m_2) = (4, 4)$ & 
\begin{tabular}{@{}l@{}}
$(\boldsymbol{\chi_1, \chi_2}) = (0.6, 0.5)$ \\
$(\boldsymbol{\chi_1, \chi_2}) = (0.3, 0.2)$ \\
$(\boldsymbol{\chi_1, \chi_2}) = (0.5, 0.05)$ \\
$(\boldsymbol{\chi_1, \chi_2}) = (0.6, 0.2)$
\end{tabular}
& 100 &
\begin{tabular}{@{}l@{}}
1677.65 \\
1374.82 \\
1395.87 \\
1511.4
\end{tabular} \\
\hline
\end{tabular}
\caption{Details of the injected parameters and corresponding ISCO frequencies.}
\label{tab:masses_spins_snr}
\end{table*} 

\section{Results and Discussions}
\label{sec:results}

\subsection{Analysing binary black hole signals in a model agnostic way}
\label{subsec:bbh}
To create BBH signals, we assume $\delta\kappa_1=\delta\kappa_2=\delta\kappa^{\rm{BH}}=0$ and $\Lambda_1=\Lambda_2=\Lambda^{\rm{BH}}=0$. As described in Table~\ref{tab:masses_spins_snr}, we choose binaries of varying masses, spins, and signal strengths. The parameter inference includes three independent recovery models. 
\begin{enumerate}
    \item a BBH model: no SIQM or tidal-deformability parameters in the recovery,  $\rm{\{\theta_{BBH}\}}$
    \item SIQM model: along with BBH parameters, only $\delta\kappa_i$ are estimated, $\rm{\{\theta_{BBH}, \delta\kappa_i\}}$ 
    \item tidal deformability model: along with BBH parameters, only $\Lambda_i$ are estimated,  $\rm{\{\theta_{BBH}, \Lambda_i\}}$
\end{enumerate}
 The recovery models are designed from a null-test perspective, meaning it does not rely on any specific alternative model, such as boson stars or other non-BH compact objects. Instead, the estimates of $\delta\kappa_i$ and $\Lambda_i$ can directly constrain the allowed parameter space of such alternative models. For instance, if the estimated value of $\delta\kappa_i$ lies within the range of $-5$ to $5$ at a given statistical credibility level, this measurement can rule out alternative models predicting $|\delta\kappa_i| > 5$ with the same level of confidence.  
Ideally, including additional parameters in the analysis, such as $\delta\kappa_i$ and $\Lambda_i$, should lead to broader posteriors for the standard BBH parameters,  $\rm{\theta_{BBH}}$, reflecting increased measurement uncertainty, even at high SNRs of around $\sim 100$. As a result, the posterior distributions of the BBH parameters may not fully capture  the injected values due to potential biases introduced by correlations with the additional parameters $\delta\kappa_i$ and $\Lambda_i$. Hence, we present the posteriors for $\delta\kappa_i$ and $\Lambda_i$ and a model comparison analysis.
 More specifically, we focus on the biases present in binary parameter inference, by measuring $\delta\kappa_i$ and  $\Lambda_i$, separately and independently, along with $\rm{\theta_{BBH}}$.

To perform this analysis, we compare the estimates of $\rm{\theta_{BBH}}$ obtained from the three different cases of recovery: a BBH model, SIQM model and tidal-deformability model.  We are interested in whether there are noticeable biases in the inferences when using the SIQM or tidal deformability models for the recovery.
 
\begin{figure*}
    \centering
   \includegraphics[width= 3. in]{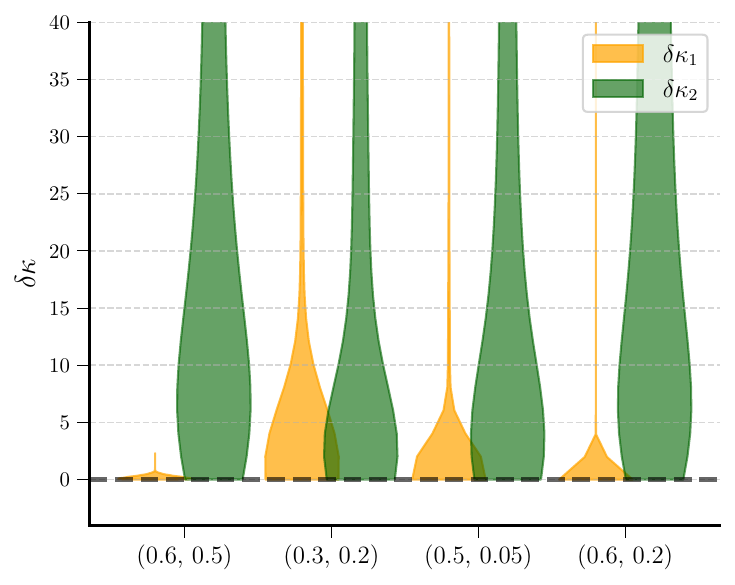}
    \includegraphics[width= 3. in]{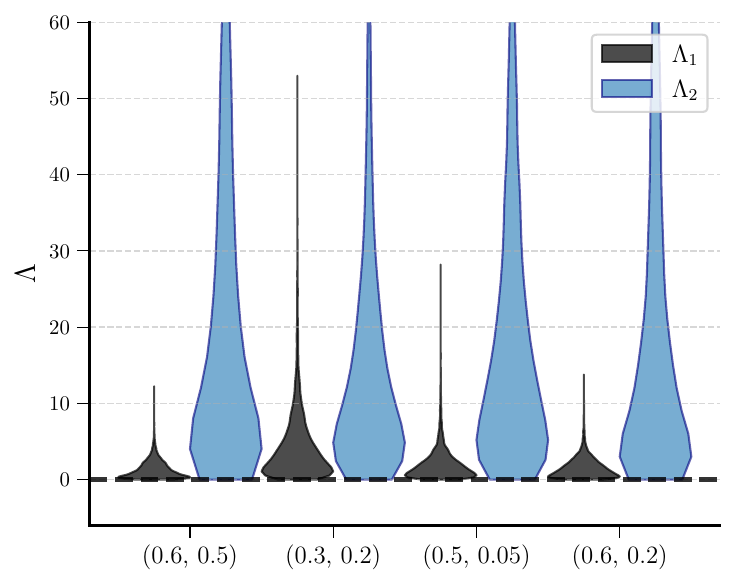}
    \caption{The posterior estimates on the SIQM parameters (left) and tidal deformability parameters (right) for a binary of masses $\rm{(5, 3)M_{\odot}}$  for BBH injections considering different spin configurations ranging $(\boldsymbol{\chi_1, \chi_2}) =(0.6, 0.5), (0.3, 0.2), (0.5, 0.05), (0.6, 0.2)$.}
    \label{fig:BBH_inj_K_L}
\end{figure*}

 In Fig.~\ref{fig:BBH_inj_K_L}, we show the posterior distributions on the testing parameters, $\delta\kappa_i$ and  $\Lambda_i$,  analyzed assuming the SIQM model and tidal deformability model, respectively. Four spin configurations are considered for a binary of masses $\rm{(5, 3)M_{\odot}}$ and signal-to-noise ratio of 50. The estimates are consistent with the BBH assumption  with no noticeable biases. As we expect, parameters corresponding to the primary BH ($\delta\kappa_1$ and  $\Lambda_1$) are better estimated compared to those of the secondary~\cite{Krishnendu:2018nqa}.  The widths of the tidal deformability posteriors also show some dependence on the spin magnitudes. This could be a consequence of the spin-dependent frequency cut-off we employ, as the duration of the signal in the band depends on the spins. As tidal deformability is a higher-order effect that is more effective at late times, its estimation may, therefore, depend on the spin magnitudes.

\begin{figure}[t]
    \includegraphics[width= 3.5 in]{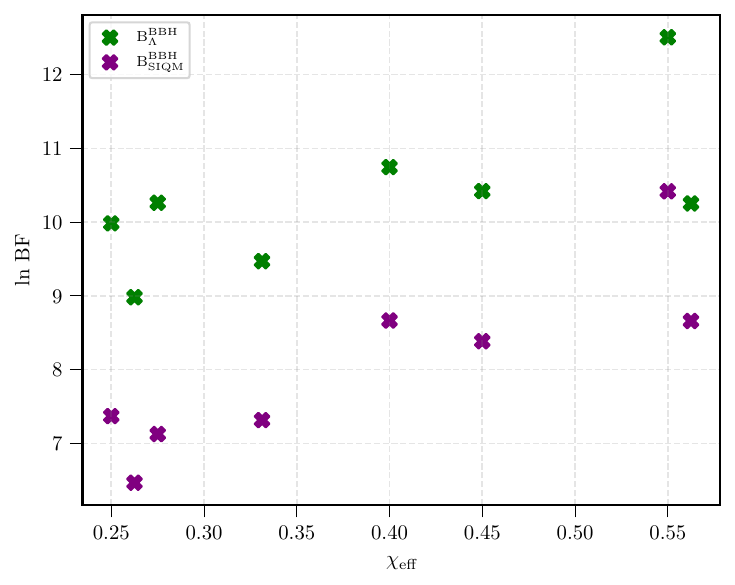}
    \caption{The Bayes factors comparing the BBH hypothesis versus a non-BBH hypothesis  for BBH injections, considering SIQM model (purple) and a tidal-deformability model (green)  for recovery. The effective spin parameter values are given on the x-axis. The positive Bayes factors suggest that the correct model is recovered in all cases, with slightly more support for the tidal-deformability model compared to the SIQM model.}
    \label{fig:BBH_inj_BFs}
\end{figure}

 Next, we compare the Bayes factors of the two non-BH models with the BBH model in Fig.~\ref{fig:BBH_inj_BFs}. The natural logarithms for the Bayes factors are plotted as a function of effective spin paramete for the BBH vs SIQM model (purple, $\rm{B^{BBH}_{SIQM}}$) and BBH vs tidal deformability model (green, $\rm{B^{BBH}_{\Lambda}}$).
The effective spin parameter contains information about the individual binary spins and masses, in the following way,
\begin{equation}
   \boldsymbol{ \chi}_{\rm{eff}}=\frac{m_1 \boldsymbol{\chi}_\mathrm{1} + m_2 \boldsymbol{\chi}_\mathrm{2}}{m_1+m_2},
\end{equation}
 The effective spin parameter represents the parameter combination that is correlated with SIQM parameters through masses and spins. The equivalent combination for the tidal deformability parameter estimation is the mass ratio.  Here, a lower BF indicates our inability to distinguish a BBH mimicker from a BBH. 

 Primarily, as the injections are BBH signals, we find the BFs always support the BBH hypothesis as expected. The tidal deformability-based test is more efficient in confirming the BBH nature for the representative systems we studied. This is due to the strong degeneracy the SIQM parameters has with spins and masses. 
However, as spins increase, the SIQM method also improves its efficiency. Although one should be cautious when comparing the effectiveness of two parametric models using Bayes factors~\cite{Chua:2019wwt}, the distinct physics represented by these two parametrization makes the comparison both meaningful and useful.

To summarize, from the BBH simulation study, we find that if the detected signal is a BBH signal,  potential biases are within our statistical errors.  In other words, given the statistical uncertainties, both SIQM and tidal deformability analyses agree with the BBH hypothesis for all the cases we considered.  This is evident from the posteriors as well as from the Bayes factors.  
We do not report the results from simultaneously estimating the SIQM and tidal-deformability parameters for these simulations, as the wider posteriors in those cases, due to additional parameters in the analysis, do not alter our conclusions. 
Next, we discuss the non-BH simulations with self-spinning boson star signatures and corresponding systematic errors when we analyze these signals with different recovery models.
%

\subsection{Analysing a boson star binary}
\label{subsec:bosonstar}
In this section, we discuss the findings from analysing a binary system with non-BH signatures,  modelled by assuming the properties of a self-spinning boson star system. We create synthetic binary signals taking into account for the SIQM and tidal deformability effects predicted for spinning boson stars. We focus on binaries with different masses and spin configurations as given in Tab.~\ref{tab:masses_spins_snr}. These synthetic signals are analyzed assuming four different recovery models: 
 \begin{enumerate}
     \item with a BBH model, $\rm{\{\theta_{BBH}\}}$
     \item SIQM-only model,  $\rm{\{\theta_{BBH}, \delta\kappa_i\}}$
     \item tidal-deformability only model,  $\rm{\{\theta_{BBH}, \Lambda_i\}}$
     \item SIQM plus tidal deformability model,  $\rm{\{\theta_{BBH}, \delta\kappa_i, \Lambda_i\}}$ 
 \end{enumerate}
 See Table~\ref{tab:injection_recovery} for more details. For the first case, the recovery model is only characterized by the BBH parameters $\{\theta_{\rm{BBH}}\}$. However, in the second, third, and fourth cases in addition to the BBH parameters, the recovery space also includes testing parameters describing the SIQM and tidal deformability effects. Namely, templates used for the second, third and fourth cases respectively contain $\{\theta_{\rm{BBH}}, \delta\kappa_1, \delta\kappa_2\}$, $\{\theta_{\rm{BBH}}, \Lambda_1, \Lambda_2\}$, $\{\theta_{\rm{BBH}}, \delta\kappa_1, \delta\kappa_2, \Lambda_1, \Lambda_2\}$ indicating the increasing complexity of our recovery model for a full non-BH inference. 

 For the simulated synthetic binary signals, we demonstrate the effect of spins, mass ratios and SNRs in determining the true nature of compact objects and the systematic biases when we use an approximate model to analyze signal versus the correct model.
\subsubsection{Effect of spin configurations}
\begin{figure*}
    \centering
    \includegraphics[width=\textwidth]{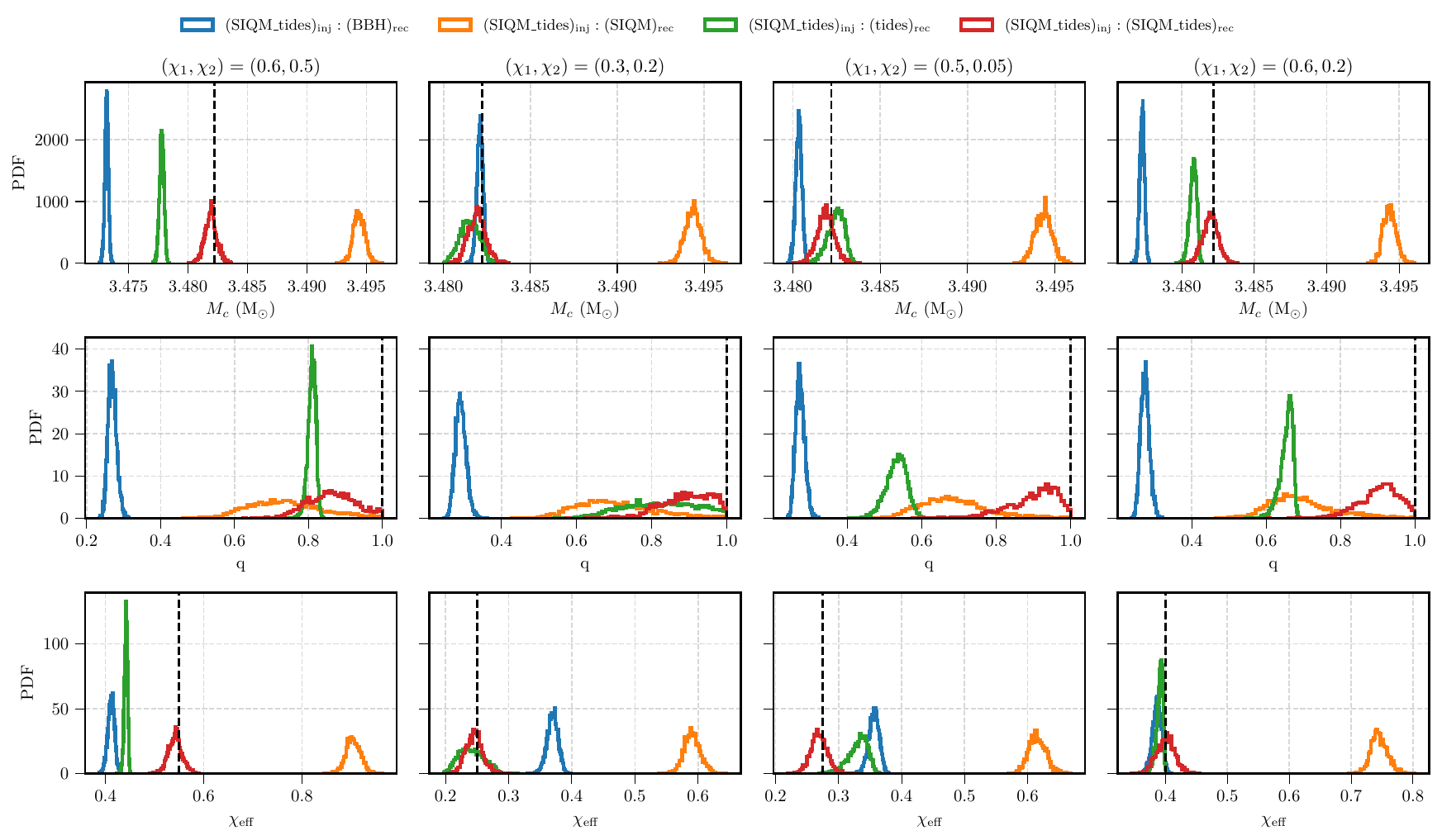}
    \caption{Posterior distributions on the chirp mass $\rm{(M_c)}$, mass ratio $\rm{(q)}$, effective spin parameter $\boldsymbol{(\chi}_{\rm eff})$  for a self-spinning boson star-like non-BBH injections with nonzero SIQM and tidal parameters (see Table~\ref{tab:injection_recovery}) for details.  Results for four different spin configurations are shown in each of the four columns: $(0.6, 0.5)$, $(0.3, 0.2)$, $(0.5, 0.05)$ and $(0.6, 0.2)$. An equal mass binary of total mass $\rm{8M_{\odot}}$ is considered at a luminosity distance of $\sim 100\rm{Mpc}$, producing a network signal-to-ratio of $100$. The blue curve in each case is obtained by considering the $\rm{(SIQM\_tides)_{inj}-(SIQM\_tides)_{rec}}$ model, orange with $\rm{(SIQM\_tides)_{inj}-(SIQM)_{rec}}$ model, green with $\rm{(SIQM\_tides)_{inj}-(tides)_{rec}}$ model and finally red is with $\rm{(SIQM\_tides)_{inj}-(BBH)_{rec}}$ model. The figure illustrates the possible bias in analyzing a non-BBH signal with an incorrect recovery model and how well the true parameters are recovered when we use the correct model. Note that the injections are represented by dashed black lines. }
    \label{fig:nBBH_siqm_tides_inj_all_4_rec_q1}
\end{figure*}
%
In Fig.~\ref{fig:nBBH_siqm_tides_inj_all_4_rec_q1}, we plot the posteriors on the chirp mass $\rm{(M_c)}$, mass ratio $\rm{(q)}$, effective spin parameter $(\boldsymbol{\chi}_{\rm eff})$ for a simulated binary with total mass $\rm{8M_{\odot}}$ and mass ratio $q=1$, considering four different spin configurations: (0.6, 0.5), (0.3, 0.2), (0.5, 0.05) and (0.6, 0.2), shown in each of the four rows. As we can see, the injected values, represented by the black dashed lines, are always recovered with the correct model, as indicated by the red histograms $\rm{(SIQM\_tides)_{inj} : (SIQM\_tides)_{rec}}$. 

However, all three other recovery models give biased estimates. For  equal mass binaries, with high nearly equal spins (0.6, 0.5), we observe the largest bias.  Comparing the SIQM model recovery $\rm{(SIQM\_tides)_{inj} : (SIQM)_{rec}}$ with the tidal deformability model recovery $\rm{(SIQM\_tides)_{inj} : (tides)_{rec}}$, the bias on $\rm{M_c}$ and $\boldsymbol{\chi_{eff}}$ is always larger for the the SIQM model, whereas the mass ratio estimate is most biased for $\rm{(SIQM\_tides)_{inj} : (tides)_{rec}}$. The spin estimates are more biased for the  $\rm{(SIQM\_tides)_{inj} : (SIQM)_{rec}}$ recovery compared to $\rm{(SIQM\_tides)_{inj} : (SIQM\_tides)_{rec}}$ recovery, as you would expect given the strong spin-dependence of the SIQM test. This bias reduces as we move to slowly spinning binaries, or at least when the secondary spin is low. Also, for  $\rm{(M_c)}$ and $\rm{(q)}$ estimation, the BBH assumption leads to the largest biased case followed by the parametric models.   Even when we use the correct model to analyze the signal, $\rm{(SIQM\_tides)_{inj} : (SIQM\_tides)_{rec}}$,  the posteriors of $q$ still shows bias, although the posteriors still encompass the injection value within $90\%$ credibility. We expect this to improve with a more accurate waveform model that includes the post-inspiral signal, additional spin effects, and higher-order harmonics.

 These results imply that if we analyze a binary boson star signal, with masses $\rm{(4,4)M_{\odot}}$, with an (incorrect) BBH model, we will identify this binary of masses $\rm{(8, 2)M_{\odot}}$,  which will likely be interpreted as a neutron star-BH binary. If we only consider the SIQM (tidal deformability) in the recovery model, these estimates would be $\rm{(4.7, 3.4)M_{\odot}}$  ($\rm{(4.4, 3.5)M_{\odot}}$) respectively. 

Hence, we conclude that the incorrect assumptions in the recovery waveform while performing tests of BH nature can lead to remarkable offsets in the measured intrinsic binary parameters and hence can lead to incorrect identification of the nature of compact objects. As a consequence, the astrophysical implications of binary mergers, such as the formation, evolution, and influence of the astrophysical environment on the dynamics, can also be inferred incorrectly.  However, including at least one of the two effects helps in reducing the bias.
%
\begin{figure*}
    \centering
    \includegraphics[width= 3. in]{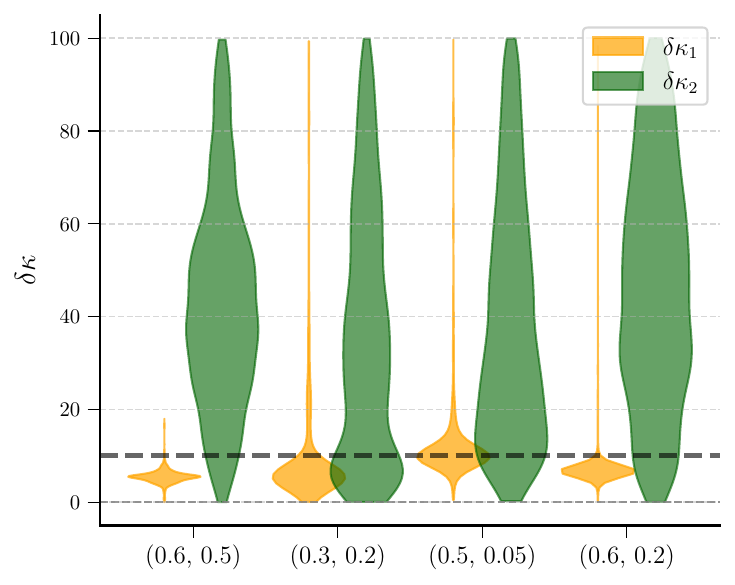}
    \includegraphics[width= 3. in]{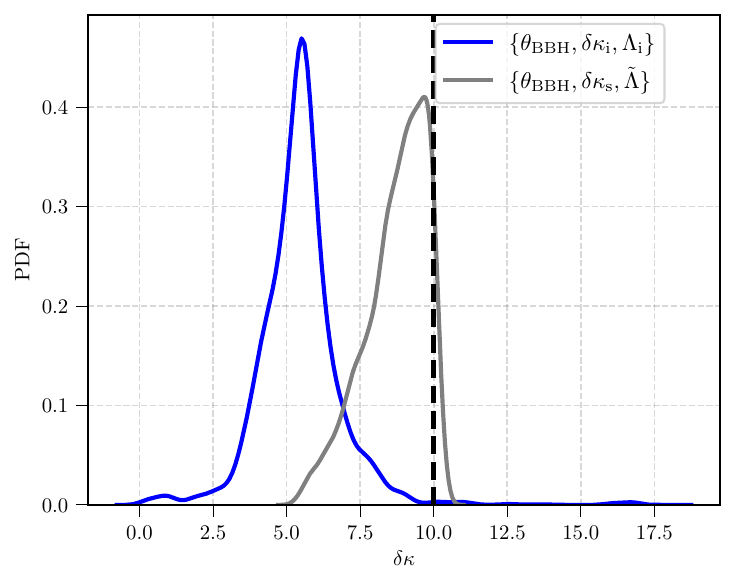}
    \caption{Left: Estimates on the $\rm{\delta\kappa_i}$ parameters for an equal mass binary with four different spin orientations, from the $\rm{(SIQM\_tides)_{inj} : (SIQM\_tides)_{rec}}$ analysis  considering non-BBH injections (see Table~\ref{tab:injection_recovery} for details). All four cases rule out the BH value, zero, strongly supporting the boson star hypothesis.  Right: Comparison of the estimates on the $\rm{\delta\kappa_s}$  parameter with $\rm{\delta\kappa_i}$ parameters from $\rm{(SIQM\_tides)_{inj} : (SIQM\_tides)_{rec}}$ analysis for a binary with spins $\rm{(0.6, 0.5)}$ and equal mass with total mass $\rm{8M_{\odot}}$. The single-parameter recovery helps reduce bias in the recovery.}
    \label{fig:nBBH_siqm_tides_inj_K_violin}
\end{figure*}

Let us now examine the $\rm{(SIQM\_tides)_{inj} : (SIQM\_tides)_{rec}}$ case in terms of recovering the non-BH feature. Figure~\ref{fig:nBBH_siqm_tides_inj_K_violin}, shows the SIQM posteriors estimated along with the tidal deformability parameter and the BBH parameters, $\rm{\{\theta_{BBH}, \delta\kappa_i, \Lambda_i\}}$. Despite the larger dimensional parameter space considered, we find that all the $\rm{\delta\kappa_1}$ estimates across the different spin configurations clearly favor a non-BH value. The most asymmetric spin combination, $(0.5, 0.05)$, provides the best estimate as indicated by the dashed black line in Fig.~\ref{fig:nBBH_siqm_tides_inj_K_violin}. Even though the $\rm{\delta\kappa_2}$ posteriors are comparatively unconstrained, there is a clear indication of less support for zero, the BH value. 
 Though this analysis considers a binary of fixed mass, the spins are varied from $(0.6, 0.5), (0.3, 0.2), (0.5, 0.05), (0.6, 0.2)$ as indicated by each column of the figure,  and the primary SIQM parameter $\delta\kappa_1$ posteriors are constrained as a function of the primary spin magnitude. This effect of spin magnitude is also reflected in the $\delta\kappa_2$ estimates. For comparable spins,  $\delta\kappa_1$ and $\delta\kappa_2$ estimates are comparable, for example, spins $\rm{(0.3, 0.2)}$. 
As evident from the same figure, though $\rm{\delta\kappa_1}$ parameter is estimated accurately and the posterior distribution is tight, but peaks at a different value compared to the injected value, $\rm{\delta\kappa_1=10}$. The reason for this bias is explained below. 

It is evident that both the  $\delta\kappa_i$ parameters are strongly correlated with each other and with other binary parameters, and their simultaneous estimation can lead to biases. 
To understand this better, we analyzed the same binary considering the  reduced parameters for the SIQM and tidal deformability models. That is, we replaced the original set of parameters with $\rm{\{\theta_{BBH}, \delta\kappa_s, \tilde{\Lambda}, \delta\Lambda}\}$. We assume $\rm{\delta\kappa_a=0}$ in this case, where $\rm{\delta\kappa_s}$ and $\rm{\delta\kappa_a}$ are the symmetric and anti-symmetric combinations of $\rm{\delta\kappa_i}$. Also, $\rm{\tilde{\Lambda}}$ and $\delta\Lambda$ are the leading and sub-leading order tidal deformability effects (see Eq.~\ref{eq:tides}). The observed bias in the $\delta\kappa_i$ estimates for $(0.6, 0.5)$  disappears if we reduce the parameter space to $\{\delta\kappa_s\}$ instead of $\{\delta\kappa_1, \delta\kappa_2\}$. In our case, since the injection assumes $\delta\kappa_1=\delta\kappa_2=10$, the symmetric combination $\delta\kappa_s=0.5(\delta\kappa_1+\delta\kappa_2)$ indeed has the same value 10, while the anti-symmetric combination $\delta\kappa_a=0.5(\delta\kappa_1-\delta\kappa_2)$ vanishes. 

The right panel of Fig.~\ref{fig:nBBH_siqm_tides_inj_K_violin} shows the posteriors on $\delta\kappa$ from two independent parameter inferences. To obtain the $\delta\kappa_s$ estimate, a reduced parameter space with only one SIQM parameter is considered, while, for the $\delta\kappa_1$ case, both $\delta\kappa_1$ and $\delta\kappa_2$ are simultaneously estimated. This is shown for a binary with spins $\rm{(0.6, 0.5)}$ and equal mass with total mass $\rm{8M_{\odot}}$. When we consider one SIQM parameter instead of two, the estimates are less biased and the true value is recovered as indicated by the dashed black line. 

We also investigated the similar features with the tidal-deformability parameter and find that employing the effective tidal deformability parameter $\tilde{\Lambda}$ provides unbiased results compared to when analysing individual tidal deformability parameters $\Lambda_1$ and $\Lambda_2$. We do not find significant correlations between the SIQM parameters and tidal-deformability parameters, though both these parameters, individually, are strongly correlated with the masses and spins of the binary. 
A more accurate full inspiral-merger-ringdown waveform model, including non-quadrupolar modes, spin-precession effects, can help break these correlations. On top of that, adding higher-order descriptions for SIQM and tidal deformability effects will also improve these estimates.  
%
\begin{figure*}
    \centering
    \includegraphics[width= 6.5 in]{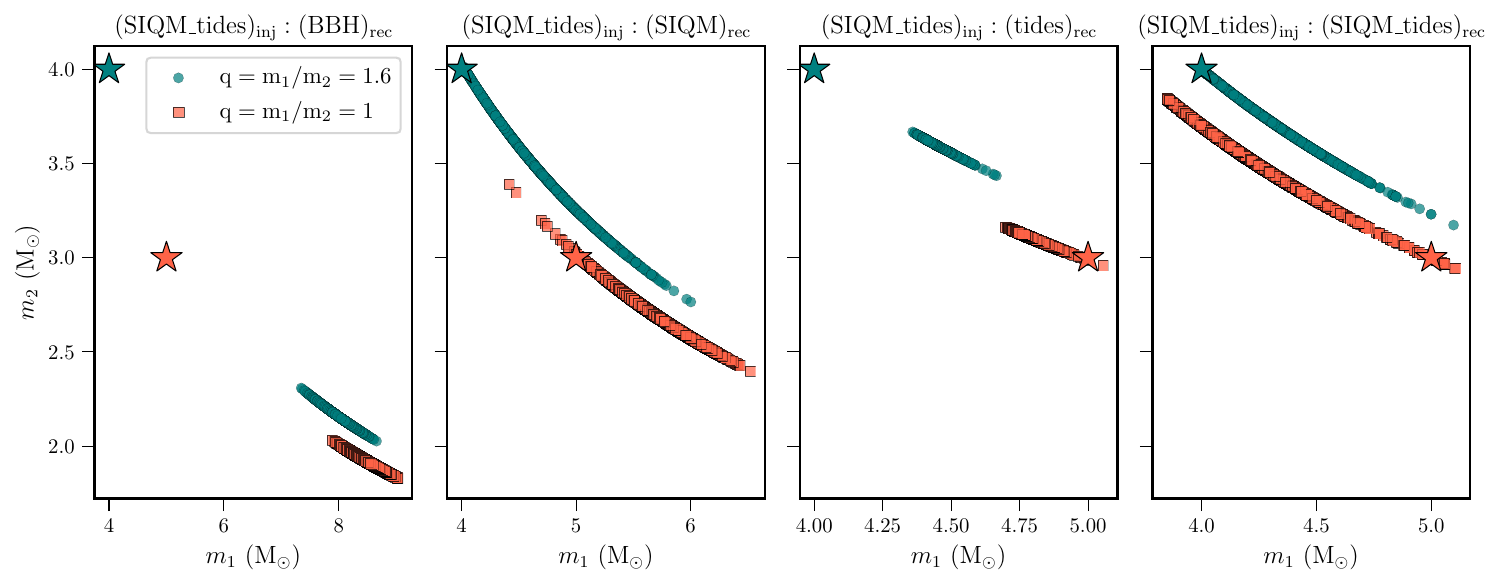}
    \caption{Biases in the inferred component masses for non-BBH injections when recovered with different models. We consider binaries with different mass ratios, $\rm{q=1}$ (solid curves) and $\rm{q=0.6}$ (dotted curves), and the spins are fixed to be $\rm{0.6, 0.5}$. The binaries produce a signal-to-noise ratio of 100 in the LIGO-Virgo detector network.  The injected values are shown as $\star$. It is evident that recovering a non-BBH signal with a BBH model can lead to severe biases in the component mass estimation.} 
    \label{fig:BS_K_L_inj_q_effect}
\end{figure*}
\subsubsection{Effect of mass ratio}
 To further study the curious effect of mass ratio bias, we plot the two-dimensional posteriors of the component masses for different mass ratios and injection/recovery models in Fig.~\ref{fig:BS_K_L_inj_q_effect}. The spins are fixed to be $\rm{0.6, 0.5}$ and binary produces an SNR of 100 in the detector network and the injected values are shown as $\star$.
For the largest biased recovery, where the recovery model is BBH template, both mass ratio estimates are comparable, whereas for the SIQM-only and tidal deformability only cases, the equal mass and unequal mass cases largely differ. Surprisingly, if we use the full model for the recovery, which includes SIQM and tidal deformability parameters, we do not see the results vary considerably with a change in the mass ratio, and we still have bias in the recovery. This bias, we believe, is due to the inspiral-only waveform we employ for the analysis.
\begin{figure*}
    \centering
    \includegraphics[width=\textwidth]{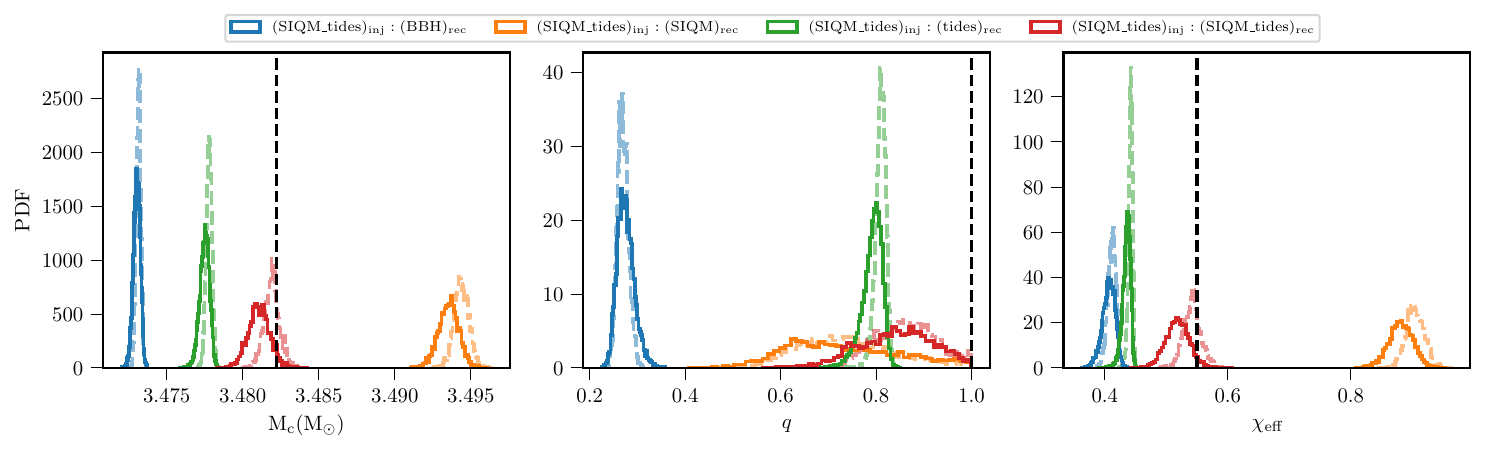}
    \caption{ Effect of SNR on the biases. For the largely spinning case with  $\boldsymbol{(\chi_1, \chi_2)}=(0.6, 0.5)$, the results assuming two different SNR values is shown, $\rm{SNR=50}$ (dashed line) and $\rm{SNR=100}$ (solid line). From left to right, the posteriors are on chirp mas, mass ratio, and effective spin parameter. As in the previous case, the total mass of the binary is chosen to be $\rm{8M_{\odot}}$ and mass ratio $\rm{q=1}$. We notice that the doubling of the SNR has a negligible effect in this case, though the posteriors are broader with SNR=50 (solid line) and are more biased compared to SNR=100 (dashed line). Note that the dashed black line represents the injected value.}
\label{fig:kappa_lambda_inj_and_rec_snr_comp_50_100}
\end{figure*}
\begin{figure*}
    \centering
    \includegraphics[width=\textwidth]{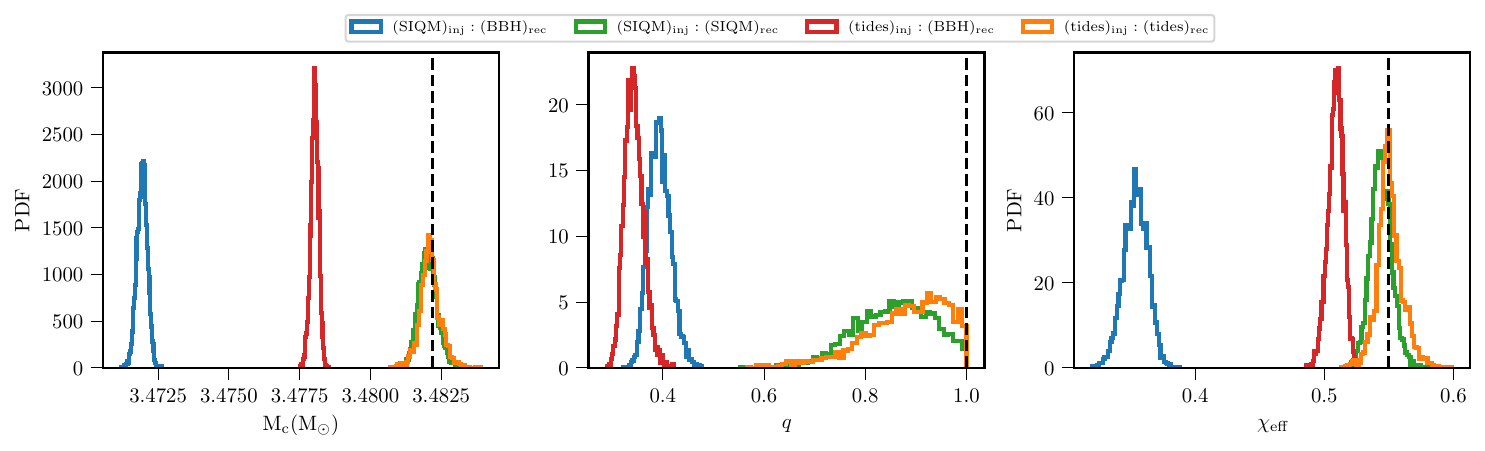}
    \includegraphics[width=\textwidth]{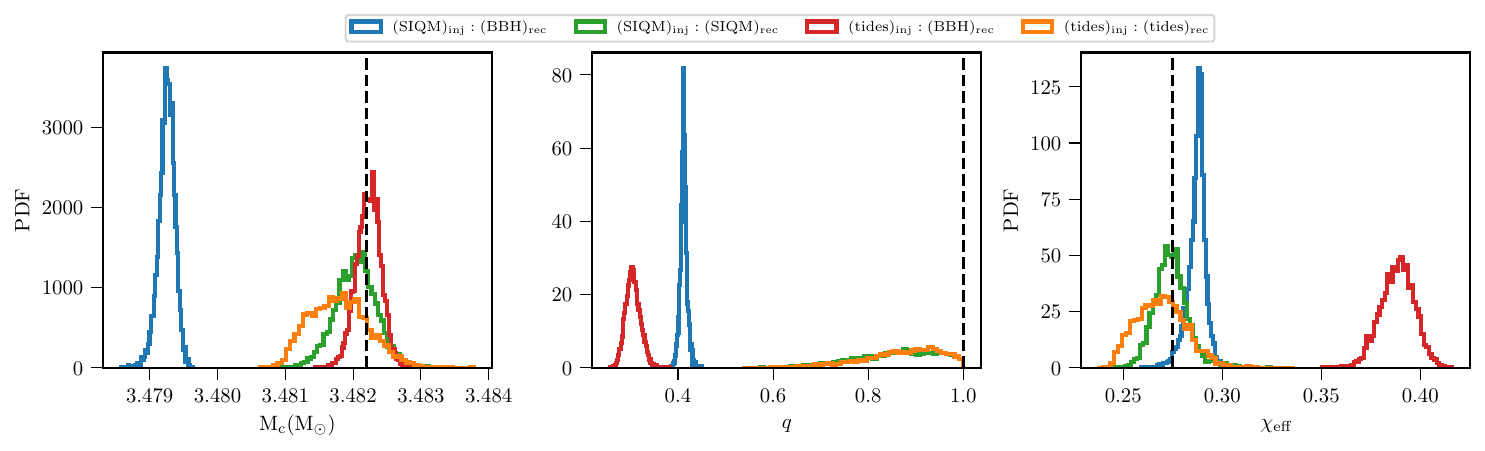}
    \caption{Posteriors on chirp mass $\rm{M_c}$, mass ratio $\rm{q}$, and the effective spin parameter $\boldsymbol{\chi}_{\rm eff}$  when the injection model considers either the SIQM or the tidal effects as a signature of non-BBH nature. The blue and green curves consider only the BBH parameters in the recovery model, $\rm{\{\theta_{BBH}\}}$ with non-BBH SIQM effects, $\rm{(SIQM)_{inj}-(BBH)_{rec}}$, and tidal deformability effects in the recovery model, $\rm{(tides)_{inj}-(BBH)_{rec}}$. On the other hand, the orange and red curves employ the correct recovery model that includes the SIQM parameters $\rm{\{\theta_{BBH}, \delta\kappa_i\}}$ and tidal deformability parameters $\rm{\{\theta_{BBH}, \Lambda_i\}}$ in the recovery model, respectively. The binary system has total mass $8\rm{M_{\odot}}$ and mass ratio $\rm{q=1}$, with spins $(0.6, 0.5)$ correspondingly to an effective spin parameter of $\boldsymbol{\chi}_{\rm eff}=0.5$ (Top Panel).  Results for a slowly spinning binary with $\boldsymbol{\chi}_{\rm eff}=0.2$ are shown in the bottom panel.}
    \label{fig:siqm_tides_sep_1}
\end{figure*}
\begin{figure*}
    \centering
    \includegraphics[width=4.5 in]{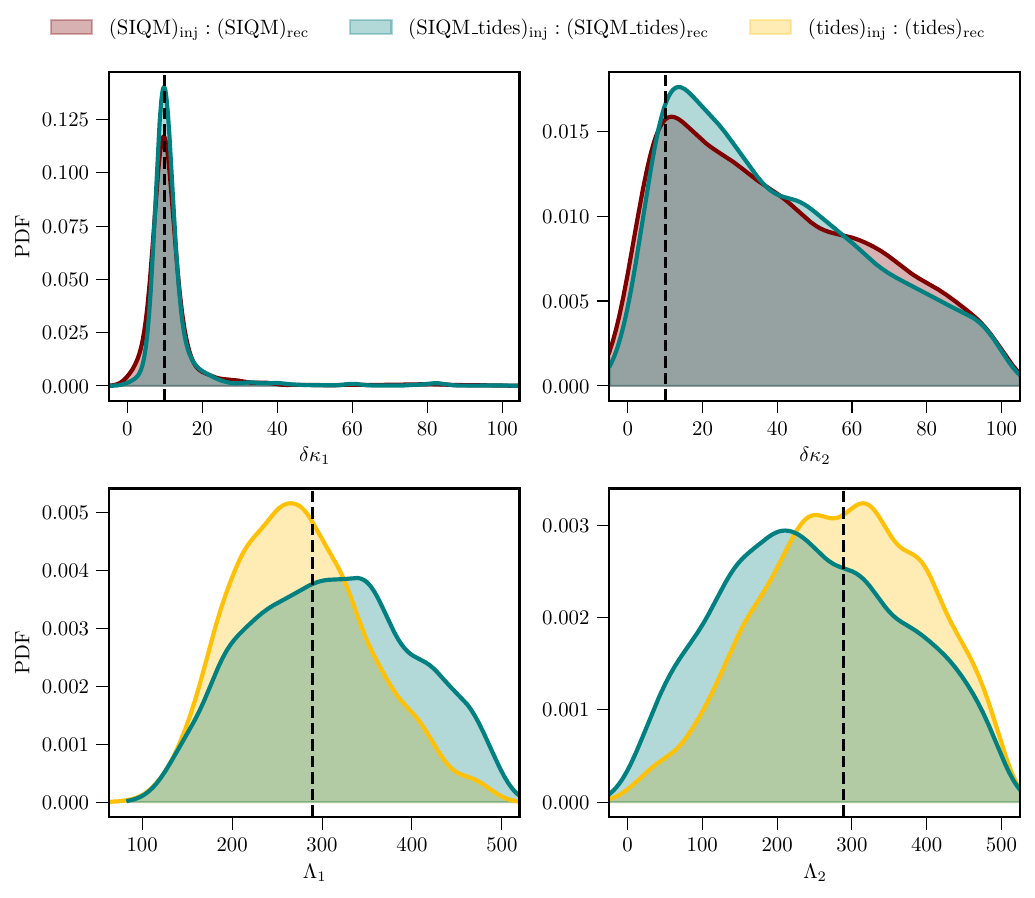}
    \caption{The comparison of $\rm{\delta\kappa_i}$ and $\Lambda_i$ measurements from a full SIQM plus tidal deformability model and from a SIQM only and tidal deformability only models. An equal mass binary with spins $\rm{(0.5, 0.05)}$, corresponding effective spin parameter $\boldsymbol{\chi}_{\rm eff}=0.5$.}
    \label{fig:siqm_tides_sep_3}
\end{figure*}
\subsubsection{Effect of SNR}
In Fig.~\ref{fig:kappa_lambda_inj_and_rec_snr_comp_50_100}, the effect of SNR in measuring the parameters of a non-BH signal is demonstrated, considering two cases, $\rm{SNR=50}$ and $\rm{SNR=100}$. We demonstrate the findings by choosing a binary of total mass $\rm{8M_{\odot}}$ and equal mass ratio, with spins $\boldsymbol{(\chi_1, \chi_2)}=(0.6, 0.5)$.  We find that the biases seen in Fig.~\ref{fig:nBBH_siqm_tides_inj_all_4_rec_q1} is not cured by doubling the SNR. The only effect of increase in the SNR is to make the posteriors narrower as well as slight shift in the recovered value towards the true value. We conclude that the biases seen are fundamental and are not corrected by simply increasing the SNR. This could potentially mean that such biases can dominate over the statistical measurement uncertainties even in the era of next-generation GW detectors.

\subsection{Simulations of boson star binary with either SIQM effect or tidal deformability effect}
\label{subsec:bosonstar2}
 Currently, the generic tests of the nature of compact objects focus on the measurement of SIQM or tidal-deformability effects separately. This is because simultaneous measurement of these parameters can lead to poor measurement due to the increased dimensionality of the parameter space. This incompleteness could potentially lead to biases in the inference.  

 To understand the possible biases in those situations, we consider two types of simulated binary signals here. First, a binary with a non-BH signature characterized by $\delta\kappa_1=\delta\kappa_2=10$ with no contributions from tidal deformability effects. On this signal, we perform a SIQM recovery where we measure both $\{\rm{\theta_{BBH}, \delta\kappa_i}\}$ $(\rm{(SIQM)_{inj}-(SIQM)_{rec}})$. Additionally, we perform a BBH recovery $(\rm{(SIQM)_{inj}-(BBH)_{rec}})$ to see the effect of neglecting SIQM effects on the recovery model, when present. Secondly, we synthesize non-BBH signals varying tidal deformability, keeping SIQM effects to zero. For this, we consider $\Lambda_1=\Lambda_2=289$ and measure $\{\rm{\theta_{BBH},\Lambda_i}\}$ $(\rm{(tides)_{inj}-(tides)_{rec}})$ and compare the estimates with the corresponding BBH recovery $(\rm{(tides)_{inj}-(BBH)_{rec}})$.  The corresponding results are shown in  the top and bottom panels of Fig.~\ref{fig:siqm_tides_sep_1}, respectively.

 As we expect, when we analyse the signal with the correct model, we see no bias. However, if we analyse the non-BBH injections with a BBH model, we end up with biases on the BBH parameters. In Fig.~\ref{fig:siqm_tides_sep_1}, the correct recovery models (orange and red curves) always include the injection value for all three parameters as indicated by the dashed black curve, whereas analyzing with a BBH model leads to significant bias in the parameter inference, especially in the highly spinning case.

 As is clear from the top panel of Fig.~\ref{fig:siqm_tides_sep_1}, the SIQM injections and BBH recovery provide the most biased $\rm{M_c}$ and $\boldsymbol{\chi_{\rm eff}}$ estimates, whereas the $\rm{q}$ bias is larger for BBH recovery with tidal deformability injections. Surprisingly, the incorrect model for inference for tidal deformability injection analysis only mildly shifts the $\boldsymbol{\chi_{\rm eff}}$ posterior. More precisely, the injection is at $0.55$ and the estimated median value of the recovered $\boldsymbol{chi_{eff}}$ is $0.54$. For both $\rm{M_c}$ and $\boldsymbol{\chi_{\rm eff}}$ parameter, neglecting SIQM effects leads to a larger bias compared to neglecting the tidal deformability effects. This is more evident in the case of $\boldsymbol{\chi_{\rm eff}}$ due to the well-known correlation between $\rm{\delta\kappa}$ and $\boldsymbol{\chi_{\rm eff}}$\cite{Krishnendu:2019tjp}.

Bottom panel of Fig.~\ref{fig:siqm_tides_sep_1} indicates that the amount of bias depends crucially on the binary parameters we consider. For instance, compared to a highly spinning binary we find a slowly spinning binary shows less bias. In this case, the tidal deformability injection recovers the correct $\rm{M_c}$ and $\boldsymbol{\chi_{\rm eff}}$ posteriors but leads to a highly biased $\rm{q}$ estimate.
It is also interesting to see that the SIQM  and tidal deformability parameter estimates agree with each other when we compare the $\rm{(SIQM)_{inj}-(SIQM)_{rec}}$ and $\rm{(SIQM\_tides)_{inj} : (SIQM\_tides)_{rec}}$ for $\delta\kappa_i$, similarly for $\rm{(SIQM)_{inj}-(tides)_{rec}}$ and $\rm{(SIQM\_tides)_{inj} : (SIQM\_tides)_{rec}}$ for the $\Lambda_i$.

In Fig.~\ref{fig:siqm_tides_sep_3}, top panel shows $\rm{\delta\kappa_1}$ and $\rm{\delta\kappa_2}$ estimates from $\rm{(SIQM)_{inj}-(SIQM)_{rec}}$ and $\rm{(SIQM\_tides)_{inj} : (SIQM\_tides)_{rec}}$ in maroon and teal colors. Similarly in the bottom panel, the $\rm{\Lambda_i}$ estimates are compared between $\rm{(SIQM)_{inj}-(tides)_{rec}}$ and $\rm{(SIQM\_tides)_{inj} : (SIQM\_tides)_{rec}}$ in yellow and teal colors. The $\rm{\Lambda_i}$ posteriors are better estimated with $\rm{(SIQM\_tides)_{inj} : (SIQM\_tides)_{rec}}$, however, there is no such trend that is seen in $\rm{\delta\kappa_i}$.
\section{Summary and Future Directions}
\label{sec:conclusion}
 Determining the true nature of low-mass compact objects—and distinguishing them from low-mass BHs—using GW observations alone is challenging. This difficulty arises from our limited understanding of the astrophysical properties, as well as the theoretical uncertainties in ruling out alternative models. Many proposed BH mimicker models exhibit features that closely resemble those of BHs~\cite{NS_BH_GW200105_GW200115, GW230529}. As a first step toward addressing this problem, we develop a Bayesian inference framework that incorporates two key physical effects—spin-induced deformations and tidal deformability measurements and assess its ability to test the BH nature of compact objects in the low mass gap with different spins. 
 
 We demonstrate the importance of simultaneously considering spin-induced deformations and tidal-deformability effects in performing tests of the nature of compact objects, focusing on compact objects with masses lying in the low mass gap. We present the method using simulated binaries with varying masses, spin magnitudes, and SNRs, employing a waveform model that simultaneously accounts for both the effects. We investigate systematic biases that arise from modelling errors when either the SIQM effect, tidal deformability, or both are omitted during the analysis. Traditional BH nature tests typically incorporate only one of these effects, either spin-induced deformations or tidal deformability, but not both. We also study how such incomplete modelling can lead to incorrect identification of the BH nature of compact objects. 

Our main finding is that the incorrectness in the model can lead to serious misidentification of the nature of binaries. For instance, a nearly equal mass highly spinning binary can be identified as a slowly spinning asymmetric binary and this could also be misidentified as a BH neutron star system. This study focuses on a specific spinning boson star model and uses an aligned-spin, inspiral-only waveform model but can be generalized to alternate models.

The waveform model used in this analysis is an inspiral-only model and does not model the post-inspiral dynamics of the binary. We justify our choice by highlighting that this study focused on low-mass binaries, hence with a large SNR in the inspiral compared to the post-inspiral. Additionally, we have no complete models available for inspiral-merger-ringdown dynamics of non-conventional compact objects, such as boson stars. The impact of higher-harmonics and spin-precession effects are higher order effects and may provide more stringent estimates on the systematic biases; however, the basic conclusions will stay. The same framework can easily be extended to more accurate waveform models, with post-inspiral dynamics, spin-precession effects, higher harmonics, and orbital-eccentricity effects. This will be studied in the future. 

In this study, we focused on the SIQM effects and the tidal deformability effects. We neglected the tidal heating, which is also an important physical effect that can be used to distinguish BHs from other exotic compact objects. Often, adding extra parameters in the recovery model leads to unconstrained estimates on all parameters and in our case, adding one extra parameter characterizing the tidal heating effects will lead to additional correlations with mass and spin parameters, along with the already existing strong correlations between SIQM and tidal deformability parameters with the binary masses and spins. Hence, we did not consider the tidal heating effects for our analysis, and they need to be explored in a separate study.  

The likelihood is truncated at the innermost stable circular frequency corresponds to a Kerr BH, and is a function of the masses and spins of the binary.  We note that this calculation does not take into account the SIQM or tidal deformability effects of the binary. For non-spinning BSs, the innermost circular frequency is calculated considering self-interacting BSs, but there are no such calculations available for spinning BS binaries~\cite{Johnson-Mcdaniel:2018cdu}.  Once the next-generation detectors become operational, we anticipate observing numerous high SNR events within the lower mass gap. Our study presents the first quantitative estimates of potential biases that may arise when one physical effect is neglected in favour of another.
\acknowledgements 
Krishnendu is supported by STFC grant ST/Y00423X/1. Frank Ohme is supported by the Max Planck Society’s Independent Research Group Grant. K. G. Arun acknowledges a grant from Infosys Foundation. Computations were carried out on the Max Planck Computing and Data Facility computing cluster Raven. We thank Poulami Dutta Roy for carefully reading the manuscript and providing us with useful comments and suggestions. This document has LIGO preprint number {\tt P2500466}. 

\newpage
\bibliographystyle{apsrev}
\bibliography{ref}

\begin{thebibliography}{108}
\expandafter\ifx\csname natexlab\endcsname\relax\def\natexlab#1{#1}\fi
\expandafter\ifx\csname bibnamefont\endcsname\relax
  \def\bibnamefont#1{#1}\fi
\expandafter\ifx\csname bibfnamefont\endcsname\relax
  \def\bibfnamefont#1{#1}\fi
\expandafter\ifx\csname citenamefont\endcsname\relax
  \def\citenamefont#1{#1}\fi
\expandafter\ifx\csname url\endcsname\relax
  \def\url#1{\texttt{#1}}\fi
\expandafter\ifx\csname urlprefix\endcsname\relax\def\urlprefix{URL }\fi
\providecommand{\bibinfo}[2]{#2}
\providecommand{\eprint}[2][]{\url{#2}}

\bibitem[{\citenamefont{Abac et~al.}(2025)}]{GWTC4}
\bibinfo{author}{\bibfnamefont{A.~G.} \bibnamefont{Abac}} \bibnamefont{et~al.}
  (\bibinfo{collaboration}{LIGO Scientific, VIRGO, KAGRA})
  (\bibinfo{year}{2025}), \eprint{2508.18082}.

\bibitem[{\citenamefont{Abbott et~al.}(2021{\natexlab{a}})}]{GWTC3-catalog}
\bibinfo{author}{\bibfnamefont{R.}~\bibnamefont{Abbott}} \bibnamefont{et~al.}
  (\bibinfo{collaboration}{LIGO Scientific, VIRGO, KAGRA})
  (\bibinfo{year}{2021}{\natexlab{a}}), \eprint{2111.03606}.

\bibitem[{\citenamefont{Abbott et~al.}(2021{\natexlab{b}})}]{GWTC-2-catalog}
\bibinfo{author}{\bibfnamefont{R.}~\bibnamefont{Abbott}} \bibnamefont{et~al.}
  (\bibinfo{collaboration}{LIGO Scientific, Virgo}), \bibinfo{journal}{Phys.
  Rev. X} \textbf{\bibinfo{volume}{11}}, \bibinfo{pages}{021053}
  (\bibinfo{year}{2021}{\natexlab{b}}), \eprint{2010.14527}.

\bibitem[{\citenamefont{Collaboration and
  Collaboration}(2024)}]{LIGO2024_gracedb}
\bibinfo{author}{\bibfnamefont{L.~S.} \bibnamefont{Collaboration}}
  \bibnamefont{and}
  \bibinfo{author}{\bibfnamefont{V.}~\bibnamefont{Collaboration}},
  \emph{\bibinfo{title}{Gravitational wave candidate event database
  (gracedb)}}, \bibinfo{howpublished}{\url{https://gracedb.ligo.org/}}
  (\bibinfo{year}{2024}).

\bibitem[{\citenamefont{Nitz et~al.}(2021)\citenamefont{Nitz, Capano, Kumar,
  Wang, Kastha, Sch{\"a}fer, Dhurkunde, and Cabero}}]{Nitz:2021uxj}
\bibinfo{author}{\bibfnamefont{A.~H.} \bibnamefont{Nitz}},
  \bibinfo{author}{\bibfnamefont{C.~D.} \bibnamefont{Capano}},
  \bibinfo{author}{\bibfnamefont{S.}~\bibnamefont{Kumar}},
  \bibinfo{author}{\bibfnamefont{Y.-F.} \bibnamefont{Wang}},
  \bibinfo{author}{\bibfnamefont{S.}~\bibnamefont{Kastha}},
  \bibinfo{author}{\bibfnamefont{M.}~\bibnamefont{Sch{\"a}fer}},
  \bibinfo{author}{\bibfnamefont{R.}~\bibnamefont{Dhurkunde}},
  \bibnamefont{and} \bibinfo{author}{\bibfnamefont{M.}~\bibnamefont{Cabero}},
  \bibinfo{journal}{Astrophys. J.} \textbf{\bibinfo{volume}{922}},
  \bibinfo{pages}{76} (\bibinfo{year}{2021}), \eprint{2105.09151}.

\bibitem[{\citenamefont{Wadekar et~al.}(2024)\citenamefont{Wadekar, Venumadhav,
  Roulet, Mehta, Zackay, Mushkin, and Zaldarriaga}}]{Wadekar:2024zdq}
\bibinfo{author}{\bibfnamefont{D.}~\bibnamefont{Wadekar}},
  \bibinfo{author}{\bibfnamefont{T.}~\bibnamefont{Venumadhav}},
  \bibinfo{author}{\bibfnamefont{J.}~\bibnamefont{Roulet}},
  \bibinfo{author}{\bibfnamefont{A.~K.} \bibnamefont{Mehta}},
  \bibinfo{author}{\bibfnamefont{B.}~\bibnamefont{Zackay}},
  \bibinfo{author}{\bibfnamefont{J.}~\bibnamefont{Mushkin}}, \bibnamefont{and}
  \bibinfo{author}{\bibfnamefont{M.}~\bibnamefont{Zaldarriaga}},
  \bibinfo{journal}{Phys. Rev. D} \textbf{\bibinfo{volume}{110}},
  \bibinfo{pages}{044063} (\bibinfo{year}{2024}), \eprint{2405.17400}.

\bibitem[{lig()}]{ligo}
\bibinfo{howpublished}{\url{http://www.ligo.caltech.edu}}.

\bibitem[{\citenamefont{Aasi et~al.}(2015)}]{AdvLIGO}
\bibinfo{author}{\bibfnamefont{J.}~\bibnamefont{Aasi}} \bibnamefont{et~al.}
  (\bibinfo{collaboration}{LIGO Scientific}), \bibinfo{journal}{Class. Quant.
  Grav.} \textbf{\bibinfo{volume}{32}}, \bibinfo{pages}{074001}
  (\bibinfo{year}{2015}), \eprint{1411.4547}.

\bibitem[{\citenamefont{Acernese et~al.}(2015)}]{AdvVirgo}
\bibinfo{author}{\bibfnamefont{F.}~\bibnamefont{Acernese}} \bibnamefont{et~al.}
  (\bibinfo{collaboration}{VIRGO}), \bibinfo{journal}{Class. Quant. Grav.}
  \textbf{\bibinfo{volume}{32}}, \bibinfo{pages}{024001}
  (\bibinfo{year}{2015}), \eprint{1408.3978}.

\bibitem[{\citenamefont{Iyer et~al.}(2011)}]{LIGO-India}
\bibinfo{author}{\bibfnamefont{B.}~\bibnamefont{Iyer}} \bibnamefont{et~al.},
  \bibinfo{type}{LIGO-India Technical Report} \bibinfo{number}{No.
  LIGO-M1100296} (\bibinfo{year}{2011}),
  \urlprefix\url{https://dcc.ligo.org/LIGO???M1100296/public/main}.

\bibitem[{\citenamefont{Evans et~al.}(2023)}]{Evans:2023euw}
\bibinfo{author}{\bibfnamefont{M.}~\bibnamefont{Evans}} \bibnamefont{et~al.}
  (\bibinfo{year}{2023}), \eprint{2306.13745}.

\bibitem[{\citenamefont{Branchesi et~al.}(2023)}]{Branchesi:2023mws}
\bibinfo{author}{\bibfnamefont{M.}~\bibnamefont{Branchesi}}
  \bibnamefont{et~al.}, \bibinfo{journal}{JCAP} \textbf{\bibinfo{volume}{07}},
  \bibinfo{pages}{068} (\bibinfo{year}{2023}), \eprint{2303.15923}.

\bibitem[{lis()}]{lisa}
\bibinfo{howpublished}{\url{http://lisa.jpl.nasa.gov}}.

\bibitem[{\citenamefont{Abbott et~al.}(2020{\natexlab{a}})\citenamefont{Abbott,
  Abbott et~al.}}]{GW190814}
\bibinfo{author}{\bibfnamefont{R.}~\bibnamefont{Abbott}},
  \bibinfo{author}{\bibfnamefont{T.~D.} \bibnamefont{Abbott}},
  \bibnamefont{et~al.}, \textbf{\bibinfo{volume}{896}}, \bibinfo{pages}{L44}
  (\bibinfo{year}{2020}{\natexlab{a}}).

\bibitem[{\citenamefont{Abac et~al.}(2024{\natexlab{a}})}]{GW230529}
\bibinfo{author}{\bibfnamefont{A.~G.} \bibnamefont{Abac}} \bibnamefont{et~al.}
  (\bibinfo{collaboration}{LIGO Scientific, KAGRA, VIRGO}),
  \bibinfo{journal}{Astrophys. J. Lett.} \textbf{\bibinfo{volume}{970}},
  \bibinfo{pages}{L34} (\bibinfo{year}{2024}{\natexlab{a}}),
  \eprint{2404.04248}.

\bibitem[{\citenamefont{Gupta et~al.}(2020)\citenamefont{Gupta, Gerosa, Arun,
  Berti, Farr, and Sathyaprakash}}]{Gupta:2019nwj}
\bibinfo{author}{\bibfnamefont{A.}~\bibnamefont{Gupta}},
  \bibinfo{author}{\bibfnamefont{D.}~\bibnamefont{Gerosa}},
  \bibinfo{author}{\bibfnamefont{K.~G.} \bibnamefont{Arun}},
  \bibinfo{author}{\bibfnamefont{E.}~\bibnamefont{Berti}},
  \bibinfo{author}{\bibfnamefont{W.~M.} \bibnamefont{Farr}}, \bibnamefont{and}
  \bibinfo{author}{\bibfnamefont{B.~S.} \bibnamefont{Sathyaprakash}},
  \bibinfo{journal}{Phys. Rev. D} \textbf{\bibinfo{volume}{101}},
  \bibinfo{pages}{103036} (\bibinfo{year}{2020}), \eprint{1909.05804}.

\bibitem[{\citenamefont{Mahapatra et~al.}(2025)\citenamefont{Mahapatra,
  Chattopadhyay, Gupta, Antonini, Favata, Sathyaprakash, and
  Arun}}]{Mahapatra:2025agb}
\bibinfo{author}{\bibfnamefont{P.}~\bibnamefont{Mahapatra}},
  \bibinfo{author}{\bibfnamefont{D.}~\bibnamefont{Chattopadhyay}},
  \bibinfo{author}{\bibfnamefont{A.}~\bibnamefont{Gupta}},
  \bibinfo{author}{\bibfnamefont{F.}~\bibnamefont{Antonini}},
  \bibinfo{author}{\bibfnamefont{M.}~\bibnamefont{Favata}},
  \bibinfo{author}{\bibfnamefont{B.~S.} \bibnamefont{Sathyaprakash}},
  \bibnamefont{and} \bibinfo{author}{\bibfnamefont{K.~G.} \bibnamefont{Arun}},
  \bibinfo{journal}{Phys. Rev. D} \textbf{\bibinfo{volume}{111}},
  \bibinfo{pages}{123030} (\bibinfo{year}{2025}), \eprint{2503.17872}.

\bibitem[{\citenamefont{Littenberg et~al.}(2015)\citenamefont{Littenberg, Farr,
  Coughlin, Kalogera, and Holz}}]{Littenberg:2015tpa}
\bibinfo{author}{\bibfnamefont{T.~B.} \bibnamefont{Littenberg}},
  \bibinfo{author}{\bibfnamefont{B.}~\bibnamefont{Farr}},
  \bibinfo{author}{\bibfnamefont{S.}~\bibnamefont{Coughlin}},
  \bibinfo{author}{\bibfnamefont{V.}~\bibnamefont{Kalogera}}, \bibnamefont{and}
  \bibinfo{author}{\bibfnamefont{D.~E.} \bibnamefont{Holz}},
  \bibinfo{journal}{Astrophys. J. Lett.} \textbf{\bibinfo{volume}{807}},
  \bibinfo{pages}{L24} (\bibinfo{year}{2015}), \eprint{1503.03179}.

\bibitem[{\citenamefont{Cotturone et~al.}(2025)\citenamefont{Cotturone, Zevin,
  and Biscoveanu}}]{Cotturone:2025jlm}
\bibinfo{author}{\bibfnamefont{J.}~\bibnamefont{Cotturone}},
  \bibinfo{author}{\bibfnamefont{M.}~\bibnamefont{Zevin}}, \bibnamefont{and}
  \bibinfo{author}{\bibfnamefont{S.}~\bibnamefont{Biscoveanu}}
  (\bibinfo{year}{2025}), \eprint{2507.01189}.

\bibitem[{\citenamefont{Johnson-Mcdaniel
  et~al.}(2020)\citenamefont{Johnson-Mcdaniel, Mukherjee, Kashyap, Ajith,
  Del~Pozzo, and Vitale}}]{Johnson-Mcdaniel:2018cdu}
\bibinfo{author}{\bibfnamefont{N.~K.} \bibnamefont{Johnson-Mcdaniel}},
  \bibinfo{author}{\bibfnamefont{A.}~\bibnamefont{Mukherjee}},
  \bibinfo{author}{\bibfnamefont{R.}~\bibnamefont{Kashyap}},
  \bibinfo{author}{\bibfnamefont{P.}~\bibnamefont{Ajith}},
  \bibinfo{author}{\bibfnamefont{W.}~\bibnamefont{Del~Pozzo}},
  \bibnamefont{and} \bibinfo{author}{\bibfnamefont{S.}~\bibnamefont{Vitale}},
  \bibinfo{journal}{Phys. Rev. D} \textbf{\bibinfo{volume}{102}},
  \bibinfo{pages}{123010} (\bibinfo{year}{2020}), \eprint{1804.08026}.

\bibitem[{\citenamefont{Krishnendu et~al.}(2017)\citenamefont{Krishnendu, Arun,
  and Mishra}}]{Krishnendu:2017shb}
\bibinfo{author}{\bibfnamefont{N.~V.} \bibnamefont{Krishnendu}},
  \bibinfo{author}{\bibfnamefont{K.~G.} \bibnamefont{Arun}}, \bibnamefont{and}
  \bibinfo{author}{\bibfnamefont{C.~K.} \bibnamefont{Mishra}},
  \bibinfo{journal}{Phys. Rev. Lett.} \textbf{\bibinfo{volume}{119}},
  \bibinfo{pages}{091101} (\bibinfo{year}{2017}), \eprint{1701.06318}.

\bibitem[{\citenamefont{Datta et~al.}(2021)\citenamefont{Datta, Phukon, and
  Bose}}]{Datta:2020gem}
\bibinfo{author}{\bibfnamefont{S.}~\bibnamefont{Datta}},
  \bibinfo{author}{\bibfnamefont{K.~S.} \bibnamefont{Phukon}},
  \bibnamefont{and} \bibinfo{author}{\bibfnamefont{S.}~\bibnamefont{Bose}},
  \bibinfo{journal}{Phys. Rev. D} \textbf{\bibinfo{volume}{104}},
  \bibinfo{pages}{084006} (\bibinfo{year}{2021}), \eprint{2004.05974}.

\bibitem[{\citenamefont{Cardoso et~al.}(2016)\citenamefont{Cardoso, Franzin,
  and Pani}}]{Cardoso2016}
\bibinfo{author}{\bibfnamefont{V.}~\bibnamefont{Cardoso}},
  \bibinfo{author}{\bibfnamefont{E.}~\bibnamefont{Franzin}}, \bibnamefont{and}
  \bibinfo{author}{\bibfnamefont{P.}~\bibnamefont{Pani}},
  \bibinfo{journal}{Phys. Rev. Lett.} \textbf{\bibinfo{volume}{116}},
  \bibinfo{pages}{171101} (\bibinfo{year}{2016}), \eprint{1602.07309}.

\bibitem[{\citenamefont{Cardoso
  et~al.}(2017{\natexlab{a}})\citenamefont{Cardoso, Franzin, Maselli, Pani, and
  Raposo}}]{CardosoTidal2017}
\bibinfo{author}{\bibfnamefont{V.}~\bibnamefont{Cardoso}},
  \bibinfo{author}{\bibfnamefont{E.}~\bibnamefont{Franzin}},
  \bibinfo{author}{\bibfnamefont{A.}~\bibnamefont{Maselli}},
  \bibinfo{author}{\bibfnamefont{P.}~\bibnamefont{Pani}}, \bibnamefont{and}
  \bibinfo{author}{\bibfnamefont{G.}~\bibnamefont{Raposo}}
  (\bibinfo{year}{2017}{\natexlab{a}}), \eprint{1701.01116}.

\bibitem[{\citenamefont{Cardoso
  et~al.}(2017{\natexlab{b}})\citenamefont{Cardoso, Franzin, Maselli, Pani, and
  Raposo}}]{Cardoso:2017cfl}
\bibinfo{author}{\bibfnamefont{V.}~\bibnamefont{Cardoso}},
  \bibinfo{author}{\bibfnamefont{E.}~\bibnamefont{Franzin}},
  \bibinfo{author}{\bibfnamefont{A.}~\bibnamefont{Maselli}},
  \bibinfo{author}{\bibfnamefont{P.}~\bibnamefont{Pani}}, \bibnamefont{and}
  \bibinfo{author}{\bibfnamefont{G.}~\bibnamefont{Raposo}},
  \bibinfo{journal}{Phys. Rev. D} \textbf{\bibinfo{volume}{95}},
  \bibinfo{pages}{084014} (\bibinfo{year}{2017}{\natexlab{b}}),
  \bibinfo{note}{[Addendum: Phys.Rev.D 95, 089901 (2017)]},
  \eprint{1701.01116}.

\bibitem[{\citenamefont{Berti and Cardoso}(2006)}]{BC06}
\bibinfo{author}{\bibfnamefont{E.}~\bibnamefont{Berti}} \bibnamefont{and}
  \bibinfo{author}{\bibfnamefont{V.}~\bibnamefont{Cardoso}},
  \bibinfo{journal}{Int. J. Mod. Phys.} \textbf{\bibinfo{volume}{D15}},
  \bibinfo{pages}{2209} (\bibinfo{year}{2006}), \eprint{gr-qc/0605101}.

\bibitem[{\citenamefont{Flanagan and Hinderer}(2008{\natexlab{a}})}]{FH08}
\bibinfo{author}{\bibfnamefont{E.~E.} \bibnamefont{Flanagan}} \bibnamefont{and}
  \bibinfo{author}{\bibfnamefont{T.}~\bibnamefont{Hinderer}},
  \bibinfo{journal}{Phys.Rev. D} \textbf{\bibinfo{volume}{77}},
  \bibinfo{pages}{021502} (\bibinfo{year}{2008}{\natexlab{a}}),
  \eprint{0709.1915}.

\bibitem[{\citenamefont{Ghosh and Hannam}(2025)}]{Ghosh:2025wex}
\bibinfo{author}{\bibfnamefont{S.}~\bibnamefont{Ghosh}} \bibnamefont{and}
  \bibinfo{author}{\bibfnamefont{M.}~\bibnamefont{Hannam}}
  (\bibinfo{year}{2025}), \eprint{2505.16380}.

\bibitem[{\citenamefont{Poisson}(1998)}]{Poisson:1997ha}
\bibinfo{author}{\bibfnamefont{E.}~\bibnamefont{Poisson}},
  \bibinfo{journal}{Phys. Rev.} \textbf{\bibinfo{volume}{D57}},
  \bibinfo{pages}{5287} (\bibinfo{year}{1998}).

\bibitem[{\citenamefont{Mishra et~al.}(2016{\natexlab{a}})\citenamefont{Mishra,
  Kela, Arun, and Faye}}]{MKAF16}
\bibinfo{author}{\bibfnamefont{C.~K.} \bibnamefont{Mishra}},
  \bibinfo{author}{\bibfnamefont{A.}~\bibnamefont{Kela}},
  \bibinfo{author}{\bibfnamefont{K.~G.} \bibnamefont{Arun}}, \bibnamefont{and}
  \bibinfo{author}{\bibfnamefont{G.}~\bibnamefont{Faye}},
  \bibinfo{journal}{Phys. Rev.} \textbf{\bibinfo{volume}{D93}},
  \bibinfo{pages}{084054} (\bibinfo{year}{2016}{\natexlab{a}}),
  \eprint{1601.05588}.

\bibitem[{\citenamefont{Bohé et~al.}(2015)\citenamefont{Bohé, Faye, and
  Marsat}}]{Bohe:2015ana}
\bibinfo{author}{\bibfnamefont{A.}~\bibnamefont{Bohé}},
  \bibinfo{author}{\bibfnamefont{G.}~\bibnamefont{Faye}}, \bibnamefont{and}
  \bibinfo{author}{\bibfnamefont{E.~K.} \bibnamefont{Marsat},
  \bibfnamefont{Sylvain acd~Porter}}, \bibinfo{journal}{Class. Quant. Grav.}
  \textbf{\bibinfo{volume}{32}}, \bibinfo{pages}{195010}
  (\bibinfo{year}{2015}), \eprint{1501.01529}.

\bibitem[{\citenamefont{Marsat}(2015{\natexlab{a}})}]{Marsat2015}
\bibinfo{author}{\bibfnamefont{S.}~\bibnamefont{Marsat}},
  \bibinfo{journal}{Class. Quant. Grav.} \textbf{\bibinfo{volume}{32}},
  \bibinfo{pages}{085008} (\bibinfo{year}{2015}{\natexlab{a}}),
  \eprint{1411.4118}.

\bibitem[{\citenamefont{Marsat}(2015{\natexlab{b}})}]{Marsat2014}
\bibinfo{author}{\bibfnamefont{S.}~\bibnamefont{Marsat}},
  \bibinfo{journal}{Class. Quant. Grav.} \textbf{\bibinfo{volume}{32}},
  \bibinfo{pages}{085008} (\bibinfo{year}{2015}{\natexlab{b}}),
  \eprint{1411.4118}.

\bibitem[{\citenamefont{Krishnendu et~al.}(2019)\citenamefont{Krishnendu,
  Saleem, Samajdar, Arun, Del~Pozzo, and Mishra}}]{Krishnendu:2019tjp}
\bibinfo{author}{\bibfnamefont{N.~V.} \bibnamefont{Krishnendu}},
  \bibinfo{author}{\bibfnamefont{M.}~\bibnamefont{Saleem}},
  \bibinfo{author}{\bibfnamefont{A.}~\bibnamefont{Samajdar}},
  \bibinfo{author}{\bibfnamefont{K.~G.} \bibnamefont{Arun}},
  \bibinfo{author}{\bibfnamefont{W.}~\bibnamefont{Del~Pozzo}},
  \bibnamefont{and} \bibinfo{author}{\bibfnamefont{C.~K.}
  \bibnamefont{Mishra}}, \bibinfo{journal}{Phys. Rev.}
  \textbf{\bibinfo{volume}{D100}}, \bibinfo{pages}{104019}
  (\bibinfo{year}{2019}), \eprint{1908.02247}.

\bibitem[{\citenamefont{Abbott et~al.}(2021{\natexlab{c}})}]{GWTC-2-TGR}
\bibinfo{author}{\bibfnamefont{R.}~\bibnamefont{Abbott}} \bibnamefont{et~al.}
  (\bibinfo{collaboration}{LIGO Scientific, Virgo}), \bibinfo{journal}{Phys.
  Rev. X} \textbf{\bibinfo{volume}{11}}, \bibinfo{pages}{021053}
  (\bibinfo{year}{2021}{\natexlab{c}}), \eprint{2010.14527}.

\bibitem[{\citenamefont{Abbott et~al.}(2021{\natexlab{d}})}]{GWTC-3-TGR}
\bibinfo{author}{\bibfnamefont{R.}~\bibnamefont{Abbott}} \bibnamefont{et~al.}
  (\bibinfo{collaboration}{LIGO Scientific, VIRGO, KAGRA})
  (\bibinfo{year}{2021}{\natexlab{d}}), \eprint{2111.03606}.

\bibitem[{\citenamefont{Ryan}(1997)}]{Ryan97}
\bibinfo{author}{\bibfnamefont{F.}~\bibnamefont{Ryan}}, \bibinfo{journal}{Phys.
  Rev. D} \textbf{\bibinfo{volume}{56}}, \bibinfo{pages}{1845}
  (\bibinfo{year}{1997}).

\bibitem[{\citenamefont{Pacilio et~al.}(2020)\citenamefont{Pacilio, Vaglio,
  Maselli, and Pani}}]{Pacilio:2020jza}
\bibinfo{author}{\bibfnamefont{C.}~\bibnamefont{Pacilio}},
  \bibinfo{author}{\bibfnamefont{M.}~\bibnamefont{Vaglio}},
  \bibinfo{author}{\bibfnamefont{A.}~\bibnamefont{Maselli}}, \bibnamefont{and}
  \bibinfo{author}{\bibfnamefont{P.}~\bibnamefont{Pani}},
  \bibinfo{journal}{Phys. Rev. D} \textbf{\bibinfo{volume}{102}},
  \bibinfo{pages}{083002} (\bibinfo{year}{2020}), \eprint{2007.05264}.

\bibitem[{\citenamefont{Pappas and Apostolatos}(2012)}]{PappasMultipole2012}
\bibinfo{author}{\bibfnamefont{G.}~\bibnamefont{Pappas}} \bibnamefont{and}
  \bibinfo{author}{\bibfnamefont{T.~A.} \bibnamefont{Apostolatos}},
  \bibinfo{journal}{Phys. Rev. Lett.} \textbf{\bibinfo{volume}{108}},
  \bibinfo{pages}{231104} (\bibinfo{year}{2012}), \eprint{1201.6067}.

\bibitem[{\citenamefont{Uchikata and Yoshida}(2016)}]{Uchikata2015}
\bibinfo{author}{\bibfnamefont{N.}~\bibnamefont{Uchikata}} \bibnamefont{and}
  \bibinfo{author}{\bibfnamefont{S.}~\bibnamefont{Yoshida}},
  \bibinfo{journal}{Class. Quant. Grav.} \textbf{\bibinfo{volume}{33}},
  \bibinfo{pages}{025005} (\bibinfo{year}{2016}), \eprint{1506.06485}.

\bibitem[{\citenamefont{Hinderer}(2008)}]{Hinderer:2007mb}
\bibinfo{author}{\bibfnamefont{T.}~\bibnamefont{Hinderer}},
  \bibinfo{journal}{Astrophys. J.} \textbf{\bibinfo{volume}{677}},
  \bibinfo{pages}{1216} (\bibinfo{year}{2008}), \bibinfo{note}{[Erratum:
  Astrophys.J. 697, 964 (2009)]}, \eprint{0711.2420}.

\bibitem[{\citenamefont{Vines et~al.}(2011{\natexlab{a}})\citenamefont{Vines,
  Flanagan, and Hinderer}}]{1PNTidal2011}
\bibinfo{author}{\bibfnamefont{J.}~\bibnamefont{Vines}},
  \bibinfo{author}{\bibfnamefont{E.~E.} \bibnamefont{Flanagan}},
  \bibnamefont{and} \bibinfo{author}{\bibfnamefont{T.}~\bibnamefont{Hinderer}},
  \bibinfo{journal}{Phys.Rev. D} \textbf{\bibinfo{volume}{83}},
  \bibinfo{pages}{084051} (\bibinfo{year}{2011}{\natexlab{a}}),
  \eprint{1101.1673}.

\bibitem[{\citenamefont{Flanagan and
  Hinderer}(2008{\natexlab{b}})}]{Flanagan:2007ix}
\bibinfo{author}{\bibfnamefont{E.~E.} \bibnamefont{Flanagan}} \bibnamefont{and}
  \bibinfo{author}{\bibfnamefont{T.}~\bibnamefont{Hinderer}},
  \bibinfo{journal}{Phys. Rev. D} \textbf{\bibinfo{volume}{77}},
  \bibinfo{pages}{021502} (\bibinfo{year}{2008}{\natexlab{b}}),
  \eprint{0709.1915}.

\bibitem[{\citenamefont{Sennett et~al.}(2017)\citenamefont{Sennett, Hinderer,
  Steinhoff, Buonanno, and Ossokine}}]{Sennett:2017etc}
\bibinfo{author}{\bibfnamefont{N.}~\bibnamefont{Sennett}},
  \bibinfo{author}{\bibfnamefont{T.}~\bibnamefont{Hinderer}},
  \bibinfo{author}{\bibfnamefont{J.}~\bibnamefont{Steinhoff}},
  \bibinfo{author}{\bibfnamefont{A.}~\bibnamefont{Buonanno}}, \bibnamefont{and}
  \bibinfo{author}{\bibfnamefont{S.}~\bibnamefont{Ossokine}},
  \bibinfo{journal}{Phys. Rev. D} \textbf{\bibinfo{volume}{96}},
  \bibinfo{pages}{024002} (\bibinfo{year}{2017}), \eprint{1704.08651}.

\bibitem[{\citenamefont{Vines et~al.}(2011{\natexlab{b}})\citenamefont{Vines,
  Flanagan, and Hinderer}}]{Vines:2011ud}
\bibinfo{author}{\bibfnamefont{J.}~\bibnamefont{Vines}},
  \bibinfo{author}{\bibfnamefont{E.~E.} \bibnamefont{Flanagan}},
  \bibnamefont{and} \bibinfo{author}{\bibfnamefont{T.}~\bibnamefont{Hinderer}},
  \bibinfo{journal}{Phys. Rev. D} \textbf{\bibinfo{volume}{83}},
  \bibinfo{pages}{084051} (\bibinfo{year}{2011}{\natexlab{b}}),
  \eprint{1101.1673}.

\bibitem[{\citenamefont{Damour et~al.}(2012)\citenamefont{Damour, Nagar, and
  Villain}}]{DNV2012}
\bibinfo{author}{\bibfnamefont{T.}~\bibnamefont{Damour}},
  \bibinfo{author}{\bibfnamefont{A.}~\bibnamefont{Nagar}}, \bibnamefont{and}
  \bibinfo{author}{\bibfnamefont{L.}~\bibnamefont{Villain}},
  \bibinfo{journal}{Phys.Rev.} \textbf{\bibinfo{volume}{D85}},
  \bibinfo{pages}{123007} (\bibinfo{year}{2012}), \eprint{1203.4352}.

\bibitem[{\citenamefont{Abbott et~al.}(2017)}]{GW170817}
\bibinfo{author}{\bibfnamefont{B.~P.} \bibnamefont{Abbott}}
  \bibnamefont{et~al.} (\bibinfo{collaboration}{Virgo, LIGO Scientific}),
  \bibinfo{journal}{Phys. Rev. Lett.} \textbf{\bibinfo{volume}{119}},
  \bibinfo{pages}{161101} (\bibinfo{year}{2017}), \eprint{1710.05832}.

\bibitem[{\citenamefont{Castro et~al.}(2022)\citenamefont{Castro, Gualtieri,
  Maselli, and Pani}}]{Castro:2022mpw}
\bibinfo{author}{\bibfnamefont{G.}~\bibnamefont{Castro}},
  \bibinfo{author}{\bibfnamefont{L.}~\bibnamefont{Gualtieri}},
  \bibinfo{author}{\bibfnamefont{A.}~\bibnamefont{Maselli}}, \bibnamefont{and}
  \bibinfo{author}{\bibfnamefont{P.}~\bibnamefont{Pani}},
  \bibinfo{journal}{Phys. Rev. D} \textbf{\bibinfo{volume}{106}},
  \bibinfo{pages}{024011} (\bibinfo{year}{2022}), \eprint{2204.12510}.

\bibitem[{\citenamefont{Abdelsalhin et~al.}(2018)\citenamefont{Abdelsalhin,
  Gualtieri, and Pani}}]{Abdelsalhin:2018reg}
\bibinfo{author}{\bibfnamefont{T.}~\bibnamefont{Abdelsalhin}},
  \bibinfo{author}{\bibfnamefont{L.}~\bibnamefont{Gualtieri}},
  \bibnamefont{and} \bibinfo{author}{\bibfnamefont{P.}~\bibnamefont{Pani}},
  \bibinfo{journal}{Phys. Rev. D} \textbf{\bibinfo{volume}{98}},
  \bibinfo{pages}{104046} (\bibinfo{year}{2018}), \eprint{1805.01487}.

\bibitem[{\citenamefont{Alvi}(2001)}]{Alvi:2001mx}
\bibinfo{author}{\bibfnamefont{K.}~\bibnamefont{Alvi}}, \bibinfo{journal}{Phys.
  Rev. D} \textbf{\bibinfo{volume}{64}}, \bibinfo{pages}{104020}
  (\bibinfo{year}{2001}), \eprint{gr-qc/0107080}.

\bibitem[{\citenamefont{Narikawa et~al.}(2021)\citenamefont{Narikawa, Uchikata,
  and Tanaka}}]{Narikawa:2021pak}
\bibinfo{author}{\bibfnamefont{T.}~\bibnamefont{Narikawa}},
  \bibinfo{author}{\bibfnamefont{N.}~\bibnamefont{Uchikata}}, \bibnamefont{and}
  \bibinfo{author}{\bibfnamefont{T.}~\bibnamefont{Tanaka}},
  \bibinfo{journal}{Phys. Rev. D} \textbf{\bibinfo{volume}{104}},
  \bibinfo{pages}{084056} (\bibinfo{year}{2021}), \eprint{2106.09193}.

\bibitem[{\citenamefont{Uchikata and Narikawa}(2021)}]{Uchikata:2021jmy}
\bibinfo{author}{\bibfnamefont{N.}~\bibnamefont{Uchikata}} \bibnamefont{and}
  \bibinfo{author}{\bibfnamefont{T.}~\bibnamefont{Narikawa}},
  \bibinfo{journal}{Phys. Rev. D} \textbf{\bibinfo{volume}{104}},
  \bibinfo{pages}{024059} (\bibinfo{year}{2021}), \eprint{2104.12968}.

\bibitem[{\citenamefont{{Ryan}}(1997)}]{Ryan97b}
\bibinfo{author}{\bibfnamefont{F.~D.} \bibnamefont{{Ryan}}},
  \bibinfo{journal}{Phys. Rev. D.} \textbf{\bibinfo{volume}{55}},
  \bibinfo{pages}{6081} (\bibinfo{year}{1997}).

\bibitem[{\citenamefont{Vaglio et~al.}(2022)\citenamefont{Vaglio, Pacilio,
  Maselli, and Pani}}]{Vaglio:2022flq}
\bibinfo{author}{\bibfnamefont{M.}~\bibnamefont{Vaglio}},
  \bibinfo{author}{\bibfnamefont{C.}~\bibnamefont{Pacilio}},
  \bibinfo{author}{\bibfnamefont{A.}~\bibnamefont{Maselli}}, \bibnamefont{and}
  \bibinfo{author}{\bibfnamefont{P.}~\bibnamefont{Pani}},
  \bibinfo{journal}{Phys. Rev. D} \textbf{\bibinfo{volume}{105}},
  \bibinfo{pages}{124020} (\bibinfo{year}{2022}), \eprint{2203.07442}.

\bibitem[{\citenamefont{Vaglio et~al.}(2023)\citenamefont{Vaglio, Pacilio,
  Maselli, and Pani}}]{Vaglio:2023lrd}
\bibinfo{author}{\bibfnamefont{M.}~\bibnamefont{Vaglio}},
  \bibinfo{author}{\bibfnamefont{C.}~\bibnamefont{Pacilio}},
  \bibinfo{author}{\bibfnamefont{A.}~\bibnamefont{Maselli}}, \bibnamefont{and}
  \bibinfo{author}{\bibfnamefont{P.}~\bibnamefont{Pani}},
  \bibinfo{journal}{Phys. Rev. D} \textbf{\bibinfo{volume}{108}},
  \bibinfo{pages}{023021} (\bibinfo{year}{2023}), \eprint{2302.13954}.

\bibitem[{\citenamefont{Abbott et~al.}(2021{\natexlab{e}})}]{GWTC-2.1-catalog}
\bibinfo{author}{\bibfnamefont{R.}~\bibnamefont{Abbott}} \bibnamefont{et~al.}
  (\bibinfo{collaboration}{LIGO Scientific, VIRGO})
  (\bibinfo{year}{2021}{\natexlab{e}}), \eprint{2108.01045}.

\bibitem[{\citenamefont{Abbott et~al.}(2020{\natexlab{b}})}]{GW190412}
\bibinfo{author}{\bibfnamefont{R.}~\bibnamefont{Abbott}} \bibnamefont{et~al.}
  (\bibinfo{collaboration}{LIGO Scientific, Virgo}), \bibinfo{journal}{Phys.
  Rev. D} \textbf{\bibinfo{volume}{102}}, \bibinfo{pages}{043015}
  (\bibinfo{year}{2020}{\natexlab{b}}), \eprint{2004.08342}.

\bibitem[{\citenamefont{Abbott et~al.}(2020{\natexlab{c}})}]{GW190425}
\bibinfo{author}{\bibfnamefont{B.~P.} \bibnamefont{Abbott}}
  \bibnamefont{et~al.} (\bibinfo{collaboration}{LIGO Scientific, Virgo}),
  \bibinfo{journal}{Astrophys. J. Lett.} \textbf{\bibinfo{volume}{892}},
  \bibinfo{pages}{L3} (\bibinfo{year}{2020}{\natexlab{c}}),
  \eprint{2001.01761}.

\bibitem[{\citenamefont{Abbott et~al.}(2020{\natexlab{d}})}]{GW190521}
\bibinfo{author}{\bibfnamefont{R.}~\bibnamefont{Abbott}} \bibnamefont{et~al.}
  (\bibinfo{collaboration}{LIGO Scientific, Virgo}), \bibinfo{journal}{Phys.
  Rev. Lett.} \textbf{\bibinfo{volume}{125}}, \bibinfo{pages}{101102}
  (\bibinfo{year}{2020}{\natexlab{d}}), \eprint{2009.01075}.

\bibitem[{\citenamefont{Hannam et~al.}(2022)}]{Hannam:2021pit}
\bibinfo{author}{\bibfnamefont{M.}~\bibnamefont{Hannam}} \bibnamefont{et~al.},
  \bibinfo{journal}{Nature} \textbf{\bibinfo{volume}{610}},
  \bibinfo{pages}{652} (\bibinfo{year}{2022}), \eprint{2112.11300}.

\bibitem[{\citenamefont{Morras et~al.}(2025)\citenamefont{Morras, Pratten, and
  Schmidt}}]{Morras:2025xfu}
\bibinfo{author}{\bibfnamefont{G.}~\bibnamefont{Morras}},
  \bibinfo{author}{\bibfnamefont{G.}~\bibnamefont{Pratten}}, \bibnamefont{and}
  \bibinfo{author}{\bibfnamefont{P.}~\bibnamefont{Schmidt}}
  (\bibinfo{year}{2025}), \eprint{2503.15393}.

\bibitem[{\citenamefont{GW230523}(2024)}]{GW230523}
\bibinfo{author}{\bibnamefont{GW230523}} (\bibinfo{collaboration}{LIGO
  Scientific, Virgo,, KAGRA, VIRGO}), \bibinfo{journal}{Astrophys. J. Lett.}
  \textbf{\bibinfo{volume}{970}}, \bibinfo{pages}{L34} (\bibinfo{year}{2024}),
  \eprint{2404.04248}.

\bibitem[{\citenamefont{Chandra et~al.}(2024)\citenamefont{Chandra, Gupta,
  Gamba, Kashyap, Chattopadhyay, Gonzalez, Bernuzzi, and
  Sathyaprakash}}]{Chandra:2024ila}
\bibinfo{author}{\bibfnamefont{K.}~\bibnamefont{Chandra}},
  \bibinfo{author}{\bibfnamefont{I.}~\bibnamefont{Gupta}},
  \bibinfo{author}{\bibfnamefont{R.}~\bibnamefont{Gamba}},
  \bibinfo{author}{\bibfnamefont{R.}~\bibnamefont{Kashyap}},
  \bibinfo{author}{\bibfnamefont{D.}~\bibnamefont{Chattopadhyay}},
  \bibinfo{author}{\bibfnamefont{A.}~\bibnamefont{Gonzalez}},
  \bibinfo{author}{\bibfnamefont{S.}~\bibnamefont{Bernuzzi}}, \bibnamefont{and}
  \bibinfo{author}{\bibfnamefont{B.~S.} \bibnamefont{Sathyaprakash}}
  (\bibinfo{year}{2024}), \eprint{2405.03841}.

\bibitem[{\citenamefont{Janquart et~al.}(2024)}]{Janquart:2024ztv}
\bibinfo{author}{\bibfnamefont{J.}~\bibnamefont{Janquart}} \bibnamefont{et~al.}
  (\bibinfo{year}{2024}), \eprint{2409.07298}.

\bibitem[{\citenamefont{van Putten et~al.}(2024)\citenamefont{van Putten,
  Aghaei~Abchouyeh, and Della~Valle}}]{vanPutten:2024ftm}
\bibinfo{author}{\bibfnamefont{M.~H. P.~M.} \bibnamefont{van Putten}},
  \bibinfo{author}{\bibfnamefont{M.}~\bibnamefont{Aghaei~Abchouyeh}},
  \bibnamefont{and}
  \bibinfo{author}{\bibfnamefont{M.}~\bibnamefont{Della~Valle}},
  \bibinfo{journal}{Astrophys. J. Lett.} \textbf{\bibinfo{volume}{972}},
  \bibinfo{pages}{L23} (\bibinfo{year}{2024}), \eprint{2408.15017}.

\bibitem[{\citenamefont{Ye et~al.}(2024)\citenamefont{Ye, Kremer, Ransom, and
  Rasio}}]{Ye:2024wqj}
\bibinfo{author}{\bibfnamefont{C.~S.} \bibnamefont{Ye}},
  \bibinfo{author}{\bibfnamefont{K.}~\bibnamefont{Kremer}},
  \bibinfo{author}{\bibfnamefont{S.~M.} \bibnamefont{Ransom}},
  \bibnamefont{and} \bibinfo{author}{\bibfnamefont{F.~A.} \bibnamefont{Rasio}}
  (\bibinfo{year}{2024}), \eprint{2408.00076}.

\bibitem[{\citenamefont{Matur et~al.}(2024)\citenamefont{Matur, Hawke, and
  Andersson}}]{Matur:2024nwi}
\bibinfo{author}{\bibfnamefont{R.}~\bibnamefont{Matur}},
  \bibinfo{author}{\bibfnamefont{I.}~\bibnamefont{Hawke}}, \bibnamefont{and}
  \bibinfo{author}{\bibfnamefont{N.}~\bibnamefont{Andersson}}
  (\bibinfo{year}{2024}), \eprint{2407.18045}.

\bibitem[{\citenamefont{Chatziioannou et~al.}(2024)\citenamefont{Chatziioannou,
  Cromartie, Gandolfi, Tews, Radice, Steiner, and
  Watts}}]{Chatziioannou:2024tjq}
\bibinfo{author}{\bibfnamefont{K.}~\bibnamefont{Chatziioannou}},
  \bibinfo{author}{\bibfnamefont{H.~T.} \bibnamefont{Cromartie}},
  \bibinfo{author}{\bibfnamefont{S.}~\bibnamefont{Gandolfi}},
  \bibinfo{author}{\bibfnamefont{I.}~\bibnamefont{Tews}},
  \bibinfo{author}{\bibfnamefont{D.}~\bibnamefont{Radice}},
  \bibinfo{author}{\bibfnamefont{A.~W.} \bibnamefont{Steiner}},
  \bibnamefont{and} \bibinfo{author}{\bibfnamefont{A.~L.} \bibnamefont{Watts}}
  (\bibinfo{year}{2024}), \eprint{2407.11153}.

\bibitem[{\citenamefont{Juli{\'e} et~al.}(2025)\citenamefont{Juli{\'e},
  Pompili, and Buonanno}}]{Julie:2024fwy}
\bibinfo{author}{\bibfnamefont{F.-L.} \bibnamefont{Juli{\'e}}},
  \bibinfo{author}{\bibfnamefont{L.}~\bibnamefont{Pompili}}, \bibnamefont{and}
  \bibinfo{author}{\bibfnamefont{A.}~\bibnamefont{Buonanno}},
  \bibinfo{journal}{Phys. Rev. D} \textbf{\bibinfo{volume}{111}},
  \bibinfo{pages}{024016} (\bibinfo{year}{2025}), \eprint{2406.13654}.

\bibitem[{\citenamefont{Gao et~al.}(2024)\citenamefont{Gao, Tang, Wang, Yan,
  and Fan}}]{Gao:2024rel}
\bibinfo{author}{\bibfnamefont{B.}~\bibnamefont{Gao}},
  \bibinfo{author}{\bibfnamefont{S.-P.} \bibnamefont{Tang}},
  \bibinfo{author}{\bibfnamefont{H.-T.} \bibnamefont{Wang}},
  \bibinfo{author}{\bibfnamefont{J.}~\bibnamefont{Yan}}, \bibnamefont{and}
  \bibinfo{author}{\bibfnamefont{Y.-Z.} \bibnamefont{Fan}},
  \bibinfo{journal}{Phys. Rev. D} \textbf{\bibinfo{volume}{110}},
  \bibinfo{pages}{044022} (\bibinfo{year}{2024}), \eprint{2405.13279}.

\bibitem[{\citenamefont{Bhattacharya et~al.}(2025)\citenamefont{Bhattacharya,
  Kapadia, and Dasgupta}}]{Bhattacharya:2025xko}
\bibinfo{author}{\bibfnamefont{S.}~\bibnamefont{Bhattacharya}},
  \bibinfo{author}{\bibfnamefont{S.}~\bibnamefont{Kapadia}}, \bibnamefont{and}
  \bibinfo{author}{\bibfnamefont{B.}~\bibnamefont{Dasgupta}}
  (\bibinfo{year}{2025}), \eprint{2507.15951}.

\bibitem[{\citenamefont{S\"anger et~al.}(2024)}]{Sanger:2024axs}
\bibinfo{author}{\bibfnamefont{E.~M.} \bibnamefont{S\"anger}}
  \bibnamefont{et~al.} (\bibinfo{year}{2024}), \eprint{2406.03568}.

\bibitem[{\citenamefont{Jedamzik}(2021)}]{PBH190814}
\bibinfo{author}{\bibfnamefont{K.}~\bibnamefont{Jedamzik}},
  \bibinfo{journal}{Phys. Rev. Lett.} \textbf{\bibinfo{volume}{126}},
  \bibinfo{pages}{051302} (\bibinfo{year}{2021}).

\bibitem[{\citenamefont{Tews et~al.}(2021)\citenamefont{Tews, Pang, Dietrich,
  Coughlin, Antier, Bulla, Heinzel, and Issa}}]{Tews:2020ylw}
\bibinfo{author}{\bibfnamefont{I.}~\bibnamefont{Tews}},
  \bibinfo{author}{\bibfnamefont{P.~T.~H.} \bibnamefont{Pang}},
  \bibinfo{author}{\bibfnamefont{T.}~\bibnamefont{Dietrich}},
  \bibinfo{author}{\bibfnamefont{M.~W.} \bibnamefont{Coughlin}},
  \bibinfo{author}{\bibfnamefont{S.}~\bibnamefont{Antier}},
  \bibinfo{author}{\bibfnamefont{M.}~\bibnamefont{Bulla}},
  \bibinfo{author}{\bibfnamefont{J.}~\bibnamefont{Heinzel}}, \bibnamefont{and}
  \bibinfo{author}{\bibfnamefont{L.}~\bibnamefont{Issa}},
  \bibinfo{journal}{Astrophys. J. Lett.} \textbf{\bibinfo{volume}{908}},
  \bibinfo{pages}{L1} (\bibinfo{year}{2021}), \eprint{2007.06057}.

\bibitem[{\citenamefont{Fasano et~al.}(2020)\citenamefont{Fasano, Wong,
  Maselli, Berti, Ferrari, and Sathyaprakash}}]{Fasano:2020eum}
\bibinfo{author}{\bibfnamefont{M.}~\bibnamefont{Fasano}},
  \bibinfo{author}{\bibfnamefont{K.~W.~K.} \bibnamefont{Wong}},
  \bibinfo{author}{\bibfnamefont{A.}~\bibnamefont{Maselli}},
  \bibinfo{author}{\bibfnamefont{E.}~\bibnamefont{Berti}},
  \bibinfo{author}{\bibfnamefont{V.}~\bibnamefont{Ferrari}}, \bibnamefont{and}
  \bibinfo{author}{\bibfnamefont{B.~S.} \bibnamefont{Sathyaprakash}},
  \bibinfo{journal}{Phys. Rev. D} \textbf{\bibinfo{volume}{102}},
  \bibinfo{pages}{023025} (\bibinfo{year}{2020}), \eprint{2005.01726}.

\bibitem[{\citenamefont{Zhang and Li}(2020)}]{Zhang:2020zsc}
\bibinfo{author}{\bibfnamefont{N.-B.} \bibnamefont{Zhang}} \bibnamefont{and}
  \bibinfo{author}{\bibfnamefont{B.-A.} \bibnamefont{Li}},
  \bibinfo{journal}{Astrophys. J.} \textbf{\bibinfo{volume}{902}},
  \bibinfo{pages}{38} (\bibinfo{year}{2020}), \eprint{2007.02513}.

\bibitem[{\citenamefont{Chen et~al.}(2020)\citenamefont{Chen, Johnson-McDaniel,
  Dietrich, and Dudi}}]{Chen:2020fzm}
\bibinfo{author}{\bibfnamefont{A.}~\bibnamefont{Chen}},
  \bibinfo{author}{\bibfnamefont{N.~K.} \bibnamefont{Johnson-McDaniel}},
  \bibinfo{author}{\bibfnamefont{T.}~\bibnamefont{Dietrich}}, \bibnamefont{and}
  \bibinfo{author}{\bibfnamefont{R.}~\bibnamefont{Dudi}},
  \bibinfo{journal}{Phys. Rev. D} \textbf{\bibinfo{volume}{101}},
  \bibinfo{pages}{103008} (\bibinfo{year}{2020}), \eprint{2001.11470}.

\bibitem[{\citenamefont{Abac
  et~al.}(2024{\natexlab{b}})}]{LIGOScientific:2024elc}
\bibinfo{author}{\bibfnamefont{A.~G.} \bibnamefont{Abac}} \bibnamefont{et~al.}
  (\bibinfo{collaboration}{LIGO Scientific, Virgo,, KAGRA, VIRGO}),
  \bibinfo{journal}{Astrophys. J. Lett.} \textbf{\bibinfo{volume}{970}},
  \bibinfo{pages}{L34} (\bibinfo{year}{2024}{\natexlab{b}}),
  \eprint{2404.04248}.

\bibitem[{\citenamefont{Colpi et~al.}(1986)\citenamefont{Colpi, Shapiro, and
  Wasserman}}]{BosonStars}
\bibinfo{author}{\bibfnamefont{M.}~\bibnamefont{Colpi}},
  \bibinfo{author}{\bibfnamefont{S.~L.} \bibnamefont{Shapiro}},
  \bibnamefont{and}
  \bibinfo{author}{\bibfnamefont{I.}~\bibnamefont{Wasserman}},
  \bibinfo{journal}{Phys. Rev. Lett.} \textbf{\bibinfo{volume}{57}},
  \bibinfo{pages}{2485} (\bibinfo{year}{1986}).

\bibitem[{\citenamefont{Uchikata et~al.}(2016)\citenamefont{Uchikata, Yoshida,
  and Pani}}]{Uchikata2016}
\bibinfo{author}{\bibfnamefont{N.}~\bibnamefont{Uchikata}},
  \bibinfo{author}{\bibfnamefont{S.}~\bibnamefont{Yoshida}}, \bibnamefont{and}
  \bibinfo{author}{\bibfnamefont{P.}~\bibnamefont{Pani}},
  \bibinfo{journal}{Phys. Rev.} \textbf{\bibinfo{volume}{D94}},
  \bibinfo{pages}{064015} (\bibinfo{year}{2016}), \eprint{1607.03593}.

\bibitem[{\citenamefont{Marsat}(2015{\natexlab{c}})}]{Marsat:2014xea}
\bibinfo{author}{\bibfnamefont{S.}~\bibnamefont{Marsat}},
  \bibinfo{journal}{Class. Quant. Grav.} \textbf{\bibinfo{volume}{32}}
  (\bibinfo{year}{2015}{\natexlab{c}}).

\bibitem[{\citenamefont{Buonanno et~al.}(2013)\citenamefont{Buonanno, Faye, and
  Hinderer}}]{BFH2012}
\bibinfo{author}{\bibfnamefont{A.}~\bibnamefont{Buonanno}},
  \bibinfo{author}{\bibfnamefont{G.}~\bibnamefont{Faye}}, \bibnamefont{and}
  \bibinfo{author}{\bibfnamefont{T.}~\bibnamefont{Hinderer}},
  \bibinfo{journal}{Phys.Rev.} \textbf{\bibinfo{volume}{D87}},
  \bibinfo{pages}{044009} (\bibinfo{year}{2013}), \eprint{1209.6349}.

\bibitem[{\citenamefont{Arun et~al.}(2009)\citenamefont{Arun, Buonanno, Faye,
  and Ochsner}}]{ABFO08}
\bibinfo{author}{\bibfnamefont{K.~G.} \bibnamefont{Arun}},
  \bibinfo{author}{\bibfnamefont{A.}~\bibnamefont{Buonanno}},
  \bibinfo{author}{\bibfnamefont{G.}~\bibnamefont{Faye}}, \bibnamefont{and}
  \bibinfo{author}{\bibfnamefont{E.}~\bibnamefont{Ochsner}},
  \bibinfo{journal}{Phys. Rev. D} \textbf{\bibinfo{volume}{79}},
  \bibinfo{pages}{104023} (\bibinfo{year}{2009}), \eprint{0810.5336}.

\bibitem[{\citenamefont{Kidder}(1995)}]{K95}
\bibinfo{author}{\bibfnamefont{L.}~\bibnamefont{Kidder}},
  \bibinfo{journal}{Phys. Rev. D} \textbf{\bibinfo{volume}{52}},
  \bibinfo{pages}{821} (\bibinfo{year}{1995}).

\bibitem[{\citenamefont{Marsat et~al.}(2013)\citenamefont{Marsat, Bohe, Faye,
  and Blanchet}}]{MBFB2012}
\bibinfo{author}{\bibfnamefont{S.}~\bibnamefont{Marsat}},
  \bibinfo{author}{\bibfnamefont{A.}~\bibnamefont{Bohe}},
  \bibinfo{author}{\bibfnamefont{G.}~\bibnamefont{Faye}}, \bibnamefont{and}
  \bibinfo{author}{\bibfnamefont{L.}~\bibnamefont{Blanchet}},
  \bibinfo{journal}{Class.Quantum Grav.} \textbf{\bibinfo{volume}{30}},
  \bibinfo{pages}{055007} (\bibinfo{year}{2013}), \eprint{arXiv:1210.4143}.

\bibitem[{\citenamefont{Bohe et~al.}(2013)\citenamefont{Bohe, Marsat, Faye, and
  Blanchet}}]{BMFB2012}
\bibinfo{author}{\bibfnamefont{A.}~\bibnamefont{Bohe}},
  \bibinfo{author}{\bibfnamefont{S.}~\bibnamefont{Marsat}},
  \bibinfo{author}{\bibfnamefont{G.}~\bibnamefont{Faye}}, \bibnamefont{and}
  \bibinfo{author}{\bibfnamefont{L.}~\bibnamefont{Blanchet}},
  \bibinfo{journal}{Class.Quant.Grav.} \textbf{\bibinfo{volume}{30}},
  \bibinfo{pages}{075017} (\bibinfo{year}{2013}), \eprint{arXiv:1212.5520}.

\bibitem[{\citenamefont{Bohé et~al.}(2013)\citenamefont{Bohé, Marsat, and
  Blanchet}}]{Bohe:2013cla}
\bibinfo{author}{\bibfnamefont{A.}~\bibnamefont{Bohé}},
  \bibinfo{author}{\bibfnamefont{S.}~\bibnamefont{Marsat}}, \bibnamefont{and}
  \bibinfo{author}{\bibfnamefont{L.}~\bibnamefont{Blanchet}},
  \bibinfo{journal}{Class. Quant. Grav.} \textbf{\bibinfo{volume}{30}},
  \bibinfo{pages}{135009} (\bibinfo{year}{2013}), \eprint{1303.7412}.

\bibitem[{\citenamefont{Marsat et~al.}(2014)\citenamefont{Marsat, Bohé,
  Blanchet, and Buonanno}}]{M3B2013}
\bibinfo{author}{\bibfnamefont{S.}~\bibnamefont{Marsat}},
  \bibinfo{author}{\bibfnamefont{A.}~\bibnamefont{Bohé}},
  \bibinfo{author}{\bibfnamefont{L.}~\bibnamefont{Blanchet}}, \bibnamefont{and}
  \bibinfo{author}{\bibfnamefont{A.}~\bibnamefont{Buonanno}},
  \bibinfo{journal}{Class.Quant.Grav.} \textbf{\bibinfo{volume}{31}},
  \bibinfo{pages}{025023} (\bibinfo{year}{2014}), \eprint{arXiv:1307.6793}.

\bibitem[{\citenamefont{Mishra et~al.}(2016{\natexlab{b}})\citenamefont{Mishra,
  Kela, Arun, and Faye}}]{Mishra:2016whh}
\bibinfo{author}{\bibfnamefont{C.~K.} \bibnamefont{Mishra}},
  \bibinfo{author}{\bibfnamefont{A.}~\bibnamefont{Kela}},
  \bibinfo{author}{\bibfnamefont{K.~G.} \bibnamefont{Arun}}, \bibnamefont{and}
  \bibinfo{author}{\bibfnamefont{G.}~\bibnamefont{Faye}},
  \bibinfo{journal}{Phys. Rev.} \textbf{\bibinfo{volume}{D93}},
  \bibinfo{pages}{084054} (\bibinfo{year}{2016}{\natexlab{b}}),
  \eprint{1601.05588}.

\bibitem[{\citenamefont{{Buonanno} et~al.}(2009)\citenamefont{{Buonanno},
  {Iyer}, {Ochsner}, {Pan}, and {Sathyaprakash}}}]{BIOPS2009}
\bibinfo{author}{\bibfnamefont{A.}~\bibnamefont{{Buonanno}}},
  \bibinfo{author}{\bibfnamefont{B.~R.} \bibnamefont{{Iyer}}},
  \bibinfo{author}{\bibfnamefont{E.}~\bibnamefont{{Ochsner}}},
  \bibinfo{author}{\bibfnamefont{Y.}~\bibnamefont{{Pan}}}, \bibnamefont{and}
  \bibinfo{author}{\bibfnamefont{B.~S.} \bibnamefont{{Sathyaprakash}}},
  \bibinfo{journal}{\prd} \textbf{\bibinfo{volume}{80}}, \bibinfo{eid}{084043}
  (\bibinfo{year}{2009}), \eprint{0907.0700}.

\bibitem[{\citenamefont{Krishnendu et~al.}(2018)\citenamefont{Krishnendu,
  Mishra, and Arun}}]{Krishnendu:2018nqa}
\bibinfo{author}{\bibfnamefont{N.~V.} \bibnamefont{Krishnendu}},
  \bibinfo{author}{\bibfnamefont{C.~K.} \bibnamefont{Mishra}},
  \bibnamefont{and} \bibinfo{author}{\bibfnamefont{K.~G.} \bibnamefont{Arun}}
  (\bibinfo{year}{2018}), \eprint{1811.00317}.

\bibitem[{\citenamefont{Lackey et~al.}(2014)\citenamefont{Lackey, Kyutoku,
  Shibata, Brady, and Friedman}}]{Lackey:2013axa}
\bibinfo{author}{\bibfnamefont{B.~D.} \bibnamefont{Lackey}},
  \bibinfo{author}{\bibfnamefont{K.}~\bibnamefont{Kyutoku}},
  \bibinfo{author}{\bibfnamefont{M.}~\bibnamefont{Shibata}},
  \bibinfo{author}{\bibfnamefont{P.~R.} \bibnamefont{Brady}}, \bibnamefont{and}
  \bibinfo{author}{\bibfnamefont{J.~L.} \bibnamefont{Friedman}},
  \bibinfo{journal}{Phys. Rev.} \textbf{\bibinfo{volume}{D89}},
  \bibinfo{pages}{043009} (\bibinfo{year}{2014}), \eprint{1303.6298}.

\bibitem[{\citenamefont{Pannarale et~al.}(2015)\citenamefont{Pannarale, Berti,
  Kyutoku, Lackey, and Shibata}}]{Pannarale:2015jka}
\bibinfo{author}{\bibfnamefont{F.}~\bibnamefont{Pannarale}},
  \bibinfo{author}{\bibfnamefont{E.}~\bibnamefont{Berti}},
  \bibinfo{author}{\bibfnamefont{K.}~\bibnamefont{Kyutoku}},
  \bibinfo{author}{\bibfnamefont{B.~D.} \bibnamefont{Lackey}},
  \bibnamefont{and} \bibinfo{author}{\bibfnamefont{M.}~\bibnamefont{Shibata}},
  \bibinfo{journal}{Phys. Rev.} \textbf{\bibinfo{volume}{D92}},
  \bibinfo{pages}{084050} (\bibinfo{year}{2015}), \eprint{1509.00512}.

\bibitem[{\citenamefont{Agathos et~al.}(2015)\citenamefont{Agathos, Meidam,
  Del~Pozzo, Li, Tompitak, Veitch, Vitale, and Van
  Den~Broeck}}]{Agathos:2015uaa}
\bibinfo{author}{\bibfnamefont{M.}~\bibnamefont{Agathos}},
  \bibinfo{author}{\bibfnamefont{J.}~\bibnamefont{Meidam}},
  \bibinfo{author}{\bibfnamefont{W.}~\bibnamefont{Del~Pozzo}},
  \bibinfo{author}{\bibfnamefont{T.~G.~F.} \bibnamefont{Li}},
  \bibinfo{author}{\bibfnamefont{M.}~\bibnamefont{Tompitak}},
  \bibinfo{author}{\bibfnamefont{J.}~\bibnamefont{Veitch}},
  \bibinfo{author}{\bibfnamefont{S.}~\bibnamefont{Vitale}}, \bibnamefont{and}
  \bibinfo{author}{\bibfnamefont{C.}~\bibnamefont{Van Den~Broeck}},
  \bibinfo{journal}{Phys. Rev.} \textbf{\bibinfo{volume}{D92}},
  \bibinfo{pages}{023012} (\bibinfo{year}{2015}), \eprint{1503.05405}.

\bibitem[{\citenamefont{Damour et~al.}(2000)\citenamefont{Damour, Iyer, and
  Sathyaprakash}}]{DIS00}
\bibinfo{author}{\bibfnamefont{T.}~\bibnamefont{Damour}},
  \bibinfo{author}{\bibfnamefont{B.~R.} \bibnamefont{Iyer}}, \bibnamefont{and}
  \bibinfo{author}{\bibfnamefont{B.~S.} \bibnamefont{Sathyaprakash}},
  \bibinfo{journal}{Phys. Rev. D} \textbf{\bibinfo{volume}{62}},
  \bibinfo{pages}{084036} (\bibinfo{year}{2000}), \eprint{gr-qc/0001023}.

\bibitem[{\citenamefont{Damour et~al.}(2001)\citenamefont{Damour, Iyer, and
  Sathyaprakash}}]{DIS01}
\bibinfo{author}{\bibfnamefont{T.}~\bibnamefont{Damour}},
  \bibinfo{author}{\bibfnamefont{B.~R.} \bibnamefont{Iyer}}, \bibnamefont{and}
  \bibinfo{author}{\bibfnamefont{B.~S.} \bibnamefont{Sathyaprakash}},
  \bibinfo{journal}{Phys. Rev. D} \textbf{\bibinfo{volume}{63}},
  \bibinfo{pages}{044023} (\bibinfo{year}{2001}), \bibinfo{note}{erratum-ibid.
  {\bf D}~72 (2005) 029902}, \eprint{gr-qc/0010009}.

\bibitem[{\citenamefont{Damour et~al.}(2002)\citenamefont{Damour, Iyer, and
  Sathyaprakash}}]{DIS02}
\bibinfo{author}{\bibfnamefont{T.}~\bibnamefont{Damour}},
  \bibinfo{author}{\bibfnamefont{B.~R.} \bibnamefont{Iyer}}, \bibnamefont{and}
  \bibinfo{author}{\bibfnamefont{B.~S.} \bibnamefont{Sathyaprakash}},
  \bibinfo{journal}{Phys. Rev. D} \textbf{\bibinfo{volume}{66}},
  \bibinfo{pages}{027502} (\bibinfo{year}{2002}),
  \bibinfo{note}{erratum-{ibid}~{\bf 66}, 027502 (2002)},
  \eprint{gr-qc/0207021}.

\bibitem[{\citenamefont{Carter}(1971)}]{Carter71}
\bibinfo{author}{\bibfnamefont{B.}~\bibnamefont{Carter}},
  \bibinfo{journal}{Phys. Rev. Lett.} \textbf{\bibinfo{volume}{26}},
  \bibinfo{pages}{331} (\bibinfo{year}{1971}),
  \urlprefix\url{http://link.aps.org/doi/10.1103/PhysRevLett.26.331}.

\bibitem[{\citenamefont{{Hansen}}(1974)}]{Hansen74}
\bibinfo{author}{\bibfnamefont{R.~O.} \bibnamefont{{Hansen}}},
  \bibinfo{journal}{Journal of Mathematical Physics}
  \textbf{\bibinfo{volume}{15}}, \bibinfo{pages}{46} (\bibinfo{year}{1974}).

\bibitem[{\citenamefont{{LIGO Scientific Collaboration}}(2018)}]{lalsuite}
\bibinfo{author}{\bibnamefont{{LIGO Scientific Collaboration}}},
  \emph{\bibinfo{title}{{LIGO} {A}lgorithm {L}ibrary - {LALS}uite}},
  \bibinfo{howpublished}{free software (GPL)} (\bibinfo{year}{2018}).

\bibitem[{\citenamefont{Ashton et~al.}(2019)}]{bilby_paper}
\bibinfo{author}{\bibfnamefont{G.}~\bibnamefont{Ashton}} \bibnamefont{et~al.},
  \bibinfo{journal}{Astrophys. J. Suppl.} \textbf{\bibinfo{volume}{241}},
  \bibinfo{pages}{27} (\bibinfo{year}{2019}), \eprint{1811.02042}.

\bibitem[{\citenamefont{Smith et~al.}(2020)\citenamefont{Smith, Ashton,
  Vajpeyi, and Talbot}}]{pbilby_paper}
\bibinfo{author}{\bibfnamefont{R.~J.~E.} \bibnamefont{Smith}},
  \bibinfo{author}{\bibfnamefont{G.}~\bibnamefont{Ashton}},
  \bibinfo{author}{\bibfnamefont{A.}~\bibnamefont{Vajpeyi}}, \bibnamefont{and}
  \bibinfo{author}{\bibfnamefont{C.}~\bibnamefont{Talbot}},
  \bibinfo{journal}{Mon. Not. Roy. Astron. Soc.}
  \textbf{\bibinfo{volume}{498}}, \bibinfo{pages}{4492} (\bibinfo{year}{2020}),
  \eprint{1909.11873}.

\bibitem[{\citenamefont{{Speagle}}(2020)}]{dynesty}
\bibinfo{author}{\bibfnamefont{J.~S.} \bibnamefont{{Speagle}}}, pp.
  \bibinfo{pages}{3132--3158} (\bibinfo{year}{2020}), \eprint{1904.02180}.

\bibitem[{\citenamefont{Favata et~al.}(2022)\citenamefont{Favata, Kim, Arun,
  Kim, and Lee}}]{Favata:2021vhw}
\bibinfo{author}{\bibfnamefont{M.}~\bibnamefont{Favata}},
  \bibinfo{author}{\bibfnamefont{C.}~\bibnamefont{Kim}},
  \bibinfo{author}{\bibfnamefont{K.~G.} \bibnamefont{Arun}},
  \bibinfo{author}{\bibfnamefont{J.}~\bibnamefont{Kim}}, \bibnamefont{and}
  \bibinfo{author}{\bibfnamefont{H.~W.} \bibnamefont{Lee}},
  \bibinfo{journal}{Phys. Rev. D} \textbf{\bibinfo{volume}{105}},
  \bibinfo{pages}{023003} (\bibinfo{year}{2022}), \eprint{2108.05861}.

\bibitem[{\citenamefont{Husa et~al.}(2016)\citenamefont{Husa, Khan, Hannam,
  P\"urrer, Ohme, Jim\'enez~Forteza, and Boh\'e}}]{Husa:2015iqa}
\bibinfo{author}{\bibfnamefont{S.}~\bibnamefont{Husa}},
  \bibinfo{author}{\bibfnamefont{S.}~\bibnamefont{Khan}},
  \bibinfo{author}{\bibfnamefont{M.}~\bibnamefont{Hannam}},
  \bibinfo{author}{\bibfnamefont{M.}~\bibnamefont{P\"urrer}},
  \bibinfo{author}{\bibfnamefont{F.}~\bibnamefont{Ohme}},
  \bibinfo{author}{\bibfnamefont{X.}~\bibnamefont{Jim\'enez~Forteza}},
  \bibnamefont{and} \bibinfo{author}{\bibfnamefont{A.}~\bibnamefont{Boh\'e}},
  \bibinfo{journal}{Phys. Rev. D} \textbf{\bibinfo{volume}{93}},
  \bibinfo{pages}{044006} (\bibinfo{year}{2016}), \eprint{1508.07250}.

\bibitem[{\citenamefont{Ori and Thorne}(2000)}]{Ori:2000zn}
\bibinfo{author}{\bibfnamefont{A.}~\bibnamefont{Ori}} \bibnamefont{and}
  \bibinfo{author}{\bibfnamefont{K.~S.} \bibnamefont{Thorne}},
  \bibinfo{journal}{Phys. Rev. D} \textbf{\bibinfo{volume}{62}},
  \bibinfo{pages}{124022} (\bibinfo{year}{2000}), \eprint{gr-qc/0003032}.

\bibitem[{\citenamefont{Chua and Vallisneri}(2020)}]{Chua:2019wwt}
\bibinfo{author}{\bibfnamefont{A.~J.~K.} \bibnamefont{Chua}} \bibnamefont{and}
  \bibinfo{author}{\bibfnamefont{M.}~\bibnamefont{Vallisneri}},
  \bibinfo{journal}{Phys. Rev. Lett.} \textbf{\bibinfo{volume}{124}},
  \bibinfo{pages}{041102} (\bibinfo{year}{2020}), \eprint{1909.05966}.

\bibitem[{\citenamefont{Abbott
  et~al.}(2021{\natexlab{f}})}]{NS_BH_GW200105_GW200115}
\bibinfo{author}{\bibfnamefont{R.}~\bibnamefont{Abbott}} \bibnamefont{et~al.}
  (\bibinfo{collaboration}{LIGO Scientific, KAGRA, VIRGO}),
  \bibinfo{journal}{Astrophys. J. Lett.} \textbf{\bibinfo{volume}{915}},
  \bibinfo{pages}{L5} (\bibinfo{year}{2021}{\natexlab{f}}),
  \eprint{2106.15163}.

\end{thebibliography}
\end{document}